\documentclass{ws-p8-50x6-00} 
\def\alt{\mathrel{\hbox{\rlap{\hbox{\lower4pt\hbox{$\sim$}}}\hbox{$<$}}}}
\def\agt{\mathrel{\hbox{\rlap{\hbox{\lower4pt\hbox{$\sim$}}}\hbox{$>$}}}}
\begin{document}
\bibliographystyle{unsrt}    

\title{An Overview of Gravitational-Wave Sources}

\author{Curt Cutler}

\address{Max Planck Institute for Gravitational Physics, 14476 Golm, Germany}

\author{Kip S. Thorne}

\address{Theoretical Astrophysics, California Institute of Technology, Pasadena, CA 91125 \\ 
\vskip12pt {\rm in consultation with}}

\author{ 
Lars Bildsten, Alessandra Buonanno, Craig Hogan,\\ 
Vassiliki 
Kalogera, Benjamin J.\ Owen, E.\ Sterl Phinney,\\ Thomas A.\ Prince, 
Frederic A.\ Rasio, Stuart L.\ Shapiro,\\ Kenneth A.\ Strain, 
Greg Ushomirsky 
and Robert V.\ Wagoner.
}

\maketitle

\abstracts{We review current best estimates of the strength and detectability
of the gravitational waves from a variety of sources, for both ground-based
and space-based detectors, and we describe the information carried
by the waves.}

\section{Preamble}\label{preamble}
\label{sec:preamble}

Major relativity conferences such as GR16 traditionally
include an overview talk on gravitational wave (GW) sources. 
Some excellent recent 
ones include those by Flanagan~\cite{flanagan_gr15}, 
Finn~\cite{finn_review}, and 
Bender et al.~\cite{hughes_etal_01}.
Such talks are always by theorists, and can be described as
basically informed speculation, since until now 
detectors sufficiently sensitive to detect GW's have not existed.
But km-scale laser interferometers (IFO's) are now coming on-line, and it
seems very likely these will detect mergers of compact binaries
within the next seven years, and possibly much sooner. There are several other
classes of sources that the ground-based IFO's might 
well detect: 
massive star collapse (supernovae
and hypernovae), rapidly rotating neutron stars, and possibly a 
GW stochastic background created in the early universe.
A space-based interferometer, LISA, is also planned (though not yet
fully funded), and could fly in $\sim 2011$. There is one type of source ---
short-period galactic binaries --- that LISA is guaranteed to observe
(at its planned sensitivity), plus a list of 
very promising candidates:
the inspiral and merger of supermassive black holes (SMBH's), the inspiral
and capture of compact
objects by SMBH's, and 
sources in the very early universe.

In this article we review the various GW sources that have been studied,
for both ground and space-based detectors, summarizing 
the best available estimates for rates and 
source strengths
and describing some of the information that may be extracted from the waves.
We will try to emphasize what is new, 
but much here is not, and indeed many sections of this paper 
have been adapted, with minor updating, from an earlier review by 
Thorne~\cite{thorne_chandra} or from a publicly available LIGO-Project
document by Thorne~\cite{thorne_LIGOII} (based on extensive input from 
the colleagues listed as consultants at the beginning of this article).

Gravitational-wave detection efforts focus on four frequency bands:
(i) The extremely low frequency (ELF) band, $10^{-15}$ to $10^{-18}$ Hz, 
in which waves will be sought via their imprint on the polarization of the
cosmic microwave background (CMB) radiation~\cite{kamio}, and the only
expected source is primordial gravitational fluctuations amplified by inflation
of the universe.  (ii) The very low frequency (VLF) band, 
$10^{-7}$ to $10^{-9}$ Hz, in which the waves are sought using high-stability 
pulsar timing~\cite{backer}, and the expected sources are processes in the 
very early universe and extremely massive binary black hole systems.
(iii) The    
low frequency (LF) band, $10^{-4}$ to 1 Hz, in which LISA will operate, and
(iv) the high frequency (HF) band,  
$1$ to $10^4$ Hz, in which earth-based detectors (interferometers and resonant-mass detectors) operate.  

In this article we shall confine attention to the
LF and HF bands, since that is where the richest variety of sources reside,
that is where the vast majority of the current experimental 
activity concentrates, and that is where the first firm detections are likely
to be made.  We shall not say much about details or current status
or plans for gravitational wave detection, as there are many other review
articles devoted to this, e.g.\ those by Kawamura~\cite{seiji}, Schutz~\cite{schutz_gr16},
Robertson~\cite{norna}, and Barish and Weiss~\cite{barish_weiss}.

In Sec.\ \ref{sec:hfsources}
we shall discuss high-frequency sources being sought by earth-based detectors,
and in Sec.\ \ref{sec:lfsources}, 
low-frequency sources that will be sought by LISA (and 
some of which are already being sought by Doppler tracking of spacecraft~\cite{doppler}).  

\section{High-Frequency Sources Sought by Earth-Based Detectors}
\label{sec:hfsources}

\subsection{Fiducial Noise Curves}
\label{subsec:LIGOII}

In discussing gravitational-wave sources, it is important to put them
in the context of detector sensitivities, both near-term and planned.
As is discussed in Kawamura's chapter in this volume~\cite{seiji},
the first-generation kilometer-scale interferometric detectors 
(LIGO, VIRGO, GEO600, TAMA300) were
being assembled at the time of GR16 and will begin their initial GW
searches about the time that our review is published.  In about 2007
these first-generation interferometers (IFOs) will be upgraded to produce
second-generation IFOs, with noise curves that can be adjusted
by adjusting the position and reflectivity of one mirror.  In parallel
with this IFO experimental effort, a network of narrow-band
resonant-mass detectors is operating and being improved~\cite{bars}.

In this review we shall find it convenient to evaluate
prospects for detection with respect to
three representative, fiducial noise curves: 
(i) that for the first-generation LIGO IFOs 
(the {\it LIGO-I} noise curve); (ii) that
for the second-generation LIGO IFOs (sometimes called {\bf advanced} LIGO
IFOs and sometimes called {\bf LIGO-II} IFOs), when they are 
operated in a ``wide-band''
mode (the {\it WB LIGO-II} noise curve); and (iii) that for second-generation
IFOs operated in a ``narrow-band'' mode that is optimized for 
sources near 600 Hz frequency (the {\it NB LIGO-II} noise curve).
Though focusing on LIGO smacks of American egocentrism, the reason is 
pragmatic: the first-generation VIRGO noise curve would largely lie between 
ithe initial LIGO (LIGO-I) curve and the advanced LIGO (LIGO-II) curves 
anyway, VIRGO does not publish an official second-generation noise
curve, and GEO600 and TAMA300 are purely first-generation instruments.

\begin{figure}
\epsfxsize=4.5in\epsfbox{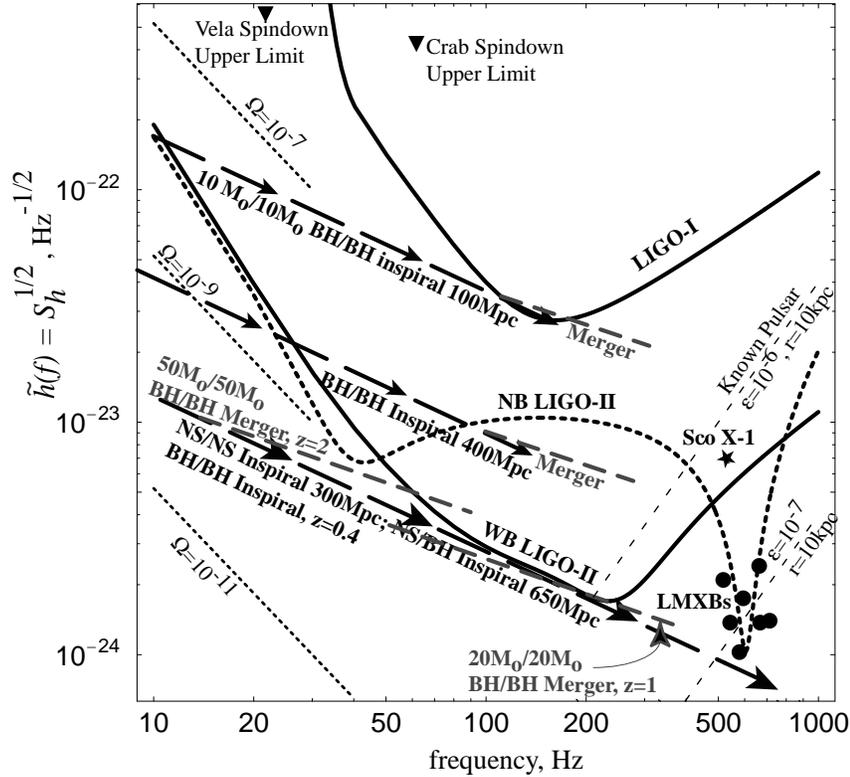}
\caption{
The noise $\tilde h(f)$ in several planned LIGO interferometers plotted 
as a function of gravity-wave frequency $f$, and compared with the estimated 
signal strengths $\tilde h_s(f)$ from various sources. The signal
strength $\tilde h_s(f)$ is defined in such a way that, 
{\it wherever a signal point or curve lies above the interferometer's
noise curve, the signal, coming from a random direction on the sky and with a
random orientation, is detectable with a false alarm probability
of less than one per cent;} see the text for greater detail and discussion.  
\label{fig:SourceSensitivity}
}
\end{figure}

Our three fiducial noise curves are shown in 
Fig.\ \ref{fig:SourceSensitivity} along with
the wave strengths for a number of HF sources to be discussed below.
The horizontal axis in the figure is GW frequency $f$; the vertical axis is
the square root of the spectral density of an
IFO's arm-length difference, i.e.\ the IFO's    
``strain per root Hertz'' $\tilde h(f) \equiv 
\sqrt{S_{\Delta L/L}(f)}$.  The (dimensionless) rms noise in a bandwidth
$\Delta f$ around frequency $f$ is given by $h_{\rm rms} = 
\tilde h(f) \sqrt{\Delta f}$.

Notice that, compared to LIGO I, the  WB LIGO-II IFOs
(i) will lower the 
amplitude noise by a factor $\sim 15$ at the frequencies of best
sensitivity $f\sim 100$--$200$ Hz, and (ii) will widen the band of high 
sensitivity
at both low frequencies (pushing it down to $\sim 20$ Hz) and high frequencies
(pushing it up to $\sim 1000$ Hz). 
The lowered WB
noise at optimal frequencies will increase the event rate for distant, 
extragalactic sources by a factor
$\sim 15^3 \simeq 3000$.  Opening up lower and higher frequency bands will 
bring us into the domains of new sources: 
colliding, massive black holes and stochastic background at
low frequencies; low-mass X-ray binaries, fast pulsars and tidal disruption 
of neutron stars by black holes at high frequencies.  Noise-curve reshaping, 
as illustrated by the NB LIGO-II IFO, 
will reduce the noise by a factor $\sim$3 to 5 relative to the WB IFO
within some 
chosen narrow frequency band $\Delta f/f \sim 0.2$ in which
targeted periodic sources (e.g. low-mass X-ray
binaries) are expected to lie.  We shall refer to this as ``narrow-band
tuning'' of the IFO. 

The expected LIGO II noise curves depend somewhat on whether the 
mirrors' substrate is made from sapphire (the preferred material)
or from fused silica (the backup).  We assume sapphire in this article.
If problems with sapphire force fused silica to be used, the WB IFOs will likely
be a few tens of percent more noisy than Fig.\ \ref{fig:SourceSensitivity}
near the noise-curve minimum, and the overall GW event rate may be
worse by about a factor 2.

Figure 1 shows, along with the noise curves, the estimated 
signal strengths $\tilde h_s(f)$ for various sources.  These signal strengths
are defined in such a way~\cite{hsDef}
that the ratio $\tilde h_s(f)/\tilde h(f) $ is equal to 
the ratio of signal $S$ to noise threshold $T$, rms averaged over source 
directions and orientations,  $\tilde h_s(f)/\tilde h(f) = \langle
S^2/T^2\rangle^{1/2}$, with the threshold being that at which the false alarm
probability is one per cent when using the best currently known, practical data
analysis algorithm.  (For broad-band sources, the algorithm is assumed to
integrate over a bandwidth equal to frequency 
and to use the output from LIGO's two 4km IFOs and one 2km
IFO~\cite{2kmToNB}, 
thereby removing all non-Gaussian noise. For periodic sources
such as spinning neutron stars, the algorithm uses data from only one 
4 km IFO, usually narrow banded, the noise again is assumed
Gaussian, and the signal is integrated for $10^7$ sec, except in cases such
as Low Mass X-Ray Binaries where
there is little gain from integrating so long.) 
This definition of $\tilde h_s(f)$ means that, 
{\it wherever a signal point or signal curve lies above the IFO
noise curve, the signal, coming from a random direction on the sky and with a
random orientation, is detectable with a false alarm probability
of less than one per cent.}  

\subsection{Overview of High-Frequency Sources}
\label{sec:HFOverview}

In this section, we shall discuss the high-frequency sources
briefly, in turn, and then in subsequent sections we shall discuss them
in greater detail, focusing on the likelihood of detection and the
science we expect to extract from detected waves. 

The three arrowed, long-dashed lines in Fig.\ 1 
represent the signal $\tilde h_s(f)$ from 
{\bf neutron-star (NS) and black-hole (BH) binaries in the last few 
minutes of their
inspiral}, assuming masses $M=1.4 M_\odot$ for each NS and $M=
10 M_\odot$ for each BH.  
These sources are best searched for by the method
of matched filters~\cite{MatchedFilters}.  Using matched filters,  
LIGO's initial IFOs
can detect NS/NS inspirals (with a 1 per cent false 
alarm probability) out to a distance of 20 Mpc [top arrowed line]; the 
wide-band advanced (LIGO-II) IFOs can do so 
15 times farther, out to 300 Mpc for NS/NS and out to 650 Mpc for NS/BH 
[bottom arrowed line].  
For BH/BH inspiral, the wide-band IFOs can see so far that cosmological 
effects are 
important.  For definiteness, throughout this proposal we assume a Hubble 
expansion rate 
$H_o = 65$ km/s/Mpc, a cold-matter density 0.4 of that required to 
close the universe $\Omega_M = 0.4$, and a vacuum energy density (cosmological
constant) 0.6 of closure $\Omega_\Lambda = 0.6$.  Then 
the wide-band IFOs can see ($10 M_\odot / 10 M_\odot$) BH/BH inspirals 
out to a cosmological redshift $z=0.4$.  The binary inspiral rates at these
advanced IFO distances are likely to be many per year; see Box~1 at the end
of this section.  The middle
arrowed line is the signal from BH/BH inspiral at 400Mpc, where
the geometric mean of the event-rate estimates for BH/BH field binaries
is three per year (third column of table in Box~1).

The {\bf tidal disruption of a NS by its BH companion} at the endpoint of NS/BH
inspiral should produce gravitational waves that carry detailed information
about the NS structure and equation of state.  The advanced IFOs may detect
these waves and extract their information; see Sec.\ \ref{sec:NSTidalDisrupt}.
This tidal disruption is a promising
candidate for the {\bf trigger of some gamma-ray bursts}, as is the final 
merger of
the two NS's in a NS/NS binary.  A gamma-burst / gravitational-wave
coincidence would be of great value in revealing the nature of gamma-burst
sources; see Sec.\ \ref{sec:GammaBurst} 

For BH/BH binaries much heavier than $\sim 10M_\odot / 10 M_\odot$, most of the
gravitational signal is likely to come from the {\bf black holes' merger and the
vibrational ringdown of the final black hole}, rather than from the 
inspiral.  Rough
estimates discussed in Sec.\ \ref{sec:BBHMerger} 
suggest that, if the holes are rapidly
spinning (within a few per cent of the fastest spin allowed, i.e.\ $a/m \agt
0.98$), then the wide-band IFOs can
see the merger waves from two $20 M_\odot$ holes out to $z=1$ and two $50
M_\odot$ holes out to $z=2$; see the down-sloping, non-arrowed, dashed lines
in Fig.\ 1.  The event rates at these distances could well be many per year,
and the waves from such mergers will carry rich physical and
astrophysical information; see Sec.\ \ref{sec:BBHMerger}.  

The triangles, star, large dots, and up-sloaping short-dashed lines 
in Fig.\ 1 represent signals from {\bf slightly
deformed, spinning neutron stars}.  The most interesting of these is a class
of objects called {\bf low-mass X-ray binaries (LMXB's)}.  
These are neutron stars
that are being torqued by accretion from a companion, but that seem to be
locked into
spin periods in the range $\sim 300$ -- 600 revolutions per second.  The most plausible
explanation for this apparent locking is that the accretion is producing an
asymmetry that radiates gravitational waves, which torque
the star's spin down at the same rate as accretion torques it 
up~\cite{bildsten,ucb}.  
Assuming this to be true, one can deduce an LMXB's wave 
strength $\tilde h_s$ from its measured X-ray flux and its spin frequency 
[with the frequency inferred, sometimes to within a few Hz but not better, 
from nearly coherent oscillations (NCOs) in type-I X-ray bursts, or 
less reliably
from frequency splittings of quasiperiodic (QPO) X-ray 
oscillations~\cite{vanderklis}.]  The spin frequency, and thence the
gravitational-wave frequency, will wander somewhat due to fluctuations in
accretion (which can be estimated by monitoring the X-ray flux) and due
to poorly known orbital parameters.  As a result, in searching for an LMXB's
waves one can only perform coherent integrations for about 20 days; thereafter,
one must stack the signals incoherently, allowing for unknown shifts of the
wave frequency~\cite{brady_creighton,papa}. 
When one uses this ``stack-slide'' method of 
data analysis, the resulting signal strengths may improve only slightly for
integration times longer than 20 days~\cite{brady_creighton}. (The
most sensitive LMXB searches will probably be with NB IFO's, with
the optimum sensitivity band re-adjusted every $\sim 20$ days to cover the
whole range $\sim 400-1000$ Hz.)

Assuming 20 days of integration
using a single 4-km IFO, the estimated signal strengths and 
frequencies for the 
strongest known LMXB's are shown by the
big dots and the star in Fig.~1.  The estimated strengths assume a
steady-state balance of accretion torque by gravitational-wave torque, and 
that the GW's are generated by density inhomogneneities 
(as opposed to NS wobble or unstable modes), so $f_{GW}$ is twice the
spin frequency~\cite{bildsten}.
For every LMXB in Fig.\ 1 except the weakest one,
the estimated rotation frequency is based on QPO splittings rather than NCOs, 
which means that about half of the frequencies
might be double these estimates: $\sim 1200$ Hz for density inhomogeneities
and $\sim 800$ Hz for r-mode sloshing, rather than $\sim 600$ Hz.
Doubling the frequency above 600 Hz reduces the emission amplitude by a 
factor 2 (at fixed
X-ray flux assuming a steady-state torque balance), and increases the amplitude
noise $S_h^{1/2}$
in an advanced NB IFO by a factor $\simeq \sqrt{2}$. As a result, at $\sim
1200$ Hz frequency, Sco X-1 would still be readily detectable in a NB IFO but
not in a WB IFO, and the
strongest of the other LMXB's would be marginal.  By contrast, if the
frequencies are
$\sim 600$ Hz or $\sim 400$ Hz and the waves are in a steady state, then  
Sco X-1 
should be very easily detected by a WB IFO, and several LMXB's should be 
detectable by narrow banding (NB).  

The advanced IFOs can perform interesting searches for waves from 
{\bf known radio pulsars},
such as the {\bf Crab and Vela} for which the current upper limits (based on
the pulsars' observed spindown rates) are shown as triangles. 
A Crab search, using coherent integrations based on the
star's observed (slightly wandering) rotation period, 
could improve the limit 
on the Crab's wave amplitude by a factor 100 and constrain the 
star's gravitational 
ellipticity to $\epsilon \alt 7\times 10^{-6}$ --- which is approaching the 
realm
of physically plausible ellipticities, $\epsilon \alt 10^{-6}$.  (We shall
discuss the physical origins of neutron-star ellipticities and their
plausible ranges in Sec.\ \ref{sec:NSspin}.)

More interesting will be 
searches
for waves from {\bf known, fast pulsars}, since 
the signal strength scales as $\tilde
h_s \propto \epsilon f^2/r$ (where $r$ is distance to the source).
The up-sloping short-dashed lines in 
Fig.\ 1 show some examples of signal strengths.
With a narrow-band IFO tuned to the vicinity of such a fast pulsar, 
the waves would be detectable when $\epsilon \agt 2\times 10^{-6} (100 {\rm
Hz}/f)^2(r/$10kpc), which is in the realm of plausible ellipticities 
for pulsars throughout our galaxy so long $f$ exceeds 200 Hz, i.e.\ the spin
frequency exceeds 100 Hz.  

Also of great interest will be searches for
{\bf previously unknown spinning neutron stars}, for
which the signal strengths $\tilde h_s$ 
will be reduced by a factor of a few to $\sim 15$ by
the lack of prior information about the frequency and its evolution and the
direction to the source (which determines the time-evolving doppler shift
produced by the earth's motion)~\cite{bccs,brady_creighton}.  
A tunable, narrow-band
IFO will be crucial to such searches.  One can search more deeply
by a factor of several, using a narrow-banded IFO that dwells on a
given frequency and its neighborhood for a few days or
weeks and then moves on to another frequency, than using a broad-interferometer
that collects signal at all frequencies simultaneously for a year.
Such searches will be in the band of physically
plausible ellipticities, for stars throughout our galaxy, if $f\agt 400$ Hz
(spin frequency above $\sim$ 200 revolutions per second).

One can search for a {\bf stochastic background} of gravitational waves by cross
correlating the outputs of LIGO's two 4 km IFOs~\cite{StochasticSearch}.  
For such a search the signal strengths $\tilde h_s(f)$ are shown in Fig.\ 1 as 
downward-sloping dotted lines, assuming a cross-correlation of 4 months of
(not necessarily contiguous) data, and isotropic waves.  The lines are labeled
by the waves' energy density
$\Omega(f)$ in a bandwidth equal to frequency and in units of the density to close
the universe, $\Omega(f) = (f dE_{\rm GW}/df)/\rho_{\rm closure}$.  Unfortunately,
the frequency of optimal {\it a priori} sensitivity, 
$f\sim 70$ Hz (where the noise
curve is parallel to the dotted $\Omega$ lines), is near the center of a
dead band for LIGO.  This dead band arises from the fact that $1/70 {\rm Hz} 
\simeq
14$ ms is about the round-trip gravity-wave travel time between the 
two LIGO sites~\cite{OverlapReduction}.  The
result is a net debilitation of the stochastic background sensitivity 
by a factor
of a few:  the initial IFOs can detect an isotropic background with
$\Omega$ down to $\sim 10^{-5}$, while advanced, wide-band IFOs can reach 
down to $\Omega \sim 5\times 10^{-9}$, and a factor 
$\sim 2$ lower than this if it is
reoptimized for low frequencies.  These are
interesting sensitivities, able to test a wide range of speculations
about the physics of the early universe and perhaps even detect waves
--- but {\it not} able to reach the level predicted by standard
inflation, $\Omega \alt 10^{-15}$.
Such a detection could have profound implications for physics and 
cosmology; see Sec.\ \ref{sec:Stochastic}.  

Other waves that advanced IFOs in LIGO will seek and may detect are those 
from {\bf the stellar core collapse that triggers supernovae} and the 
{\bf boiling of the nascent neutron star} at the endpoint of that collapse 
(Sec.\ \ref{sec:Supernovae}), {\bf accretion induced 
collapse of white dwarfs} (Sec.\ \ref{sec:Supernovae}), 
and {\bf totally unknown sources}. 

We now give a more detailed discussion of each of the high-frequency 
sources, focusing especially on event rate estimates and 
the information that the waves 
may bring.  Box 1 gives a brief
summary of the various sources and their detectability by
LIGO I and II (WB and NB).

\begin{table}[p]
\hskip -0.9in\fbox{
\begin{minipage}[c]{6.25truein}
\centerline{{\bf Box 1}}
\centerline{{\bf Brief Summary of Detection Capabilities of Advanced LIGO
(LIGO-II) Interferometers}}
\medskip

$\bullet$ {\bf Inspiral of NS/NS, NS/BH and BH/BH Binaries:}  The table below~\cite{vkalogera,2kmToNB} shows
estimated rates ${\cal R}_{\rm gal}$ in our galaxy 
(with masses $\sim 1.4 M_\odot$ for NS and 
$\sim 10 M_\odot$ for BH), the distances
${\cal D}_{\rm I}$ and ${\cal D}_{\rm WB}$ to which initial IFOs
and advanced WB IFOs can detect them, and corresponding
estimates of detection rates 
${\cal R}_{\rm I}$ and ${\cal R}_{\rm WB}$; Sec.\ \ref{sec:nsnsinspiral}.

\vskip5pt
\begin{tabular}{lllll}
\hline
\hline
\null & NS/NS & NS/BH & BH/BH in field & BH/BH in clusters\\ 
\\
${\cal R}_{\rm gal}$, yr$^{-1}$ & $10^{-6}$--$5\times 10^{-4}$ & $\alt 
10^{-7}$--$10^{-4}$ & $\alt10^{-7}$--$10^{-5}$ & $\sim 10^{-6}$--$10^{-5}$ \\ 
$D_{\rm I}$ & 20 Mpc & 43 Mpc & $100$ & $100$ \\
${\cal R}_{\rm I}$, yr$^{-1}$ & $3\times 10^{-4}$ -- 0.3 & $\alt 4\times
10^{-4}$ -- 0.6 &  $\alt 4\times 10^{-3}$ -- 0.6 & $\sim 0.04$ -- 0.6 \\
$D_{\rm WB}$ & 300 Mpc & 650 Mpc & $z=0.4$ & $z=0.4$ \\
${\cal R}_{\rm WB}$, yr$^{-1}$ & 1 -- 800 & $\alt 1$ -- 1500 & $\alt 30$ 
-- 4000 & $\sim 300$ -- 4000 \\  
\hline
\hline
\end{tabular}
\vskip5pt

$\bullet$ {\bf Tidal disruption of NS by BH in NS/BH binaries:}  
First crude estimates 
suggest WB IFOs can measure onset of disruption
at 140Mpc well enough to deduce the NS radius to 15\% accuracy
(compared to current uncertainties of a factor $\sim 2$); see table
above for rates; Sec.\ \ref{sec:NSTidalDisrupt}.

$\bullet$ {\bf BH/BH merger and ringdown:}  Rough estimates suggest 
detectability, by WB IFOs 
out to the cosmological distances shown in Fig.\ 2(b); rates
for BH/BH total mass $\sim 20 M_\odot$ are in table above; rates for
much larger masses are unknown; Sec.\ \ref{sec:BBHMerger}.

$\bullet$ {\bf Low-Mass X-Ray Binaries:}  If accretion's spin-up torque on NS 
due is counterbalanced by gravitational-wave-emission torque, then
WB IFOs can detect Sco X-1, and NB IFOs can detect $\sim 6$ other known
LMXB's; Sec.\ \ref{sec:NSspin}.

$\bullet$ {\bf Fast, Known Spinning NS's (Pulsars with pulse frequency
above 100 Hz):} Detectable by a advanced NB IFO in 3 months'
integration time, if NS ellipticity is $\epsilon \agt 2\times 10^{-8}(1000{\rm
Hz}/f)^2(r/10{\rm kpc})$, where $f$ is gravity wave frequency (twice the pulsar
frequency) and $r$ is  distance; actual ellipticities are unknown, but 
plausible range is $\epsilon \alt 10^{-6}$; Sec.\ \ref{sec:NSspin}.

$\bullet$ {\bf Fast, Unknown Spinning NS's:}  Unknown frequency wandering and
doppler shifts degrade the detectable ellipticity $\epsilon$ by a factor of
a few to $\sim 15$, so detection with a NB IFO requires
$\epsilon \agt (0.6$ to $3)\times 10^{-5}(100{\rm Hz}/f)^2(r/10{\rm kpc})$; 
Secs.\  \ref{sec:HFOverview} and \ref{sec:NSspin}.

$\bullet$ {\bf Centrifugally Hung-Up Proto Neutron Stars in White-Dwarf
Accretion-Induced Collapse and in Supernovae:}  Dynamics of
star very poorly understood; if instability deforms star into tumbling bar, may
be detectable by WB IFOs to $\sim 20$ Mpc (the Virgo Cluster), and possibly
farther; event rates uncertain but could be enough for detection; Sec.\ 
\ref{sec:Supernovae}.

$\bullet$ {\bf Convection of Supernova Core:} May be detectable by WB IFOs, 
via correlations with neutrinos, for supernovae in our Galaxy and possibly
Magellanic Clouds; Sec.\ \ref{sec:Supernovae}.

$\bullet$ {\bf Gamma Ray Bursts:} If triggered by NS/BH mergers, a few per 
year could be detectable by WB IFOs; if none are seen individually, statistical
studies could nevertheless confirm gravity-wave emission by the gamma-burst
triggers; Sec.\ \ref{sec:GammaBurst}. 

$\bullet$ {\bf Stochastic Background:} Detectable by cross correlating Hanford
and Livingston 4km detector outputs, if $\Omega = ($gravitational-wave 
energy in $\Delta f \sim
f \sim 40$ Hz$)\agt 5\times 10^{-9}$; there are many possible sources of
such waves in very early universe, all very speculative; Sec.\ 
\ref{sec:Stochastic}.
\end{minipage}
}
\end{table}

\subsection{Inspiraling NS and BH Binaries with $M_{BH} \alt 10M_\odot$}
\label{sec:nsnsinspiral}
As we have discussed, wide-band advanced-LIGO (LIGO-II) IFOs
can detect the waves from NS/NS inspirals out to 300Mpc, NS/BH out
to 650 Mpc, and BH/BH out to $z=0.4$.  The event rates out to these distances
can be estimated from observational data in our own galaxy~\cite{vkalogera}, 
and an
extrapolation out through the universe based on the density of massive stars
(which can be deduced by several different 
methods)~\cite{vkalogera,extrapolation1,extrap3}. 
The resulting rates (as compiled and evaluated by V.\ Kalogera~\cite{vkalogera})
are quite uncertain but very promising; see Box 1.

For NS/NS binaries, the event rate in our galaxy is constrained by the results
of radio astronomers' searches for binary pulsars that will merge, due to
gravitational radiation emission, in less than the age of the universe, and by
other aspects of pulsar searches~\cite{vkalogera}.  The resulting
constraints, 
$ 10^{-6}$/yr $\alt {\cal R}_{\rm gal} \alt 5\times 10^{-4}$/yr, 
extrapolate to a
NS/NS event rate for advanced IFOs between $1$/yr and 
800/yr.
Searches in our galaxy for NS/BH binaries in which the NS is a pulsar have
failed to find any as yet, so we must turn to much less reliable estimates
based on ``population synthesis'' (simulations of the evolution of a 
population of progenitor binary systems to determine the number that make NS/BH
binaries compact enough to merge in less than the universe's age).  
Population synthesis
gives a NS/BH event rate in our galaxy in the range $\sim 10^{-7}$/yr $\alt
{\cal R}_{\rm gal} \alt 10^{-4}$/yr~\cite{kalogera1}, though it is 
possible the rate could be even less than this (hence the $\alt 10^{-7}$/yr in
Box 1).  Extrapolating 
out into the universe, Kalogera finds a NS/BH rate
in WB advanced IFOs between $\alt 1$/yr and about 1500/yr.  
A similar analysis for
BH/BH binaries based on population synthesis (third column of the table
in Box 1) gives a
rate between $\alt 30$/yr and $\sim 4000$/yr
~\cite{CosmologicalExtrapolation}.  That population synthesis estimate 
ignores the likely
role of globular clusters and other types of dense star clusters as 
``machines'' for making BH/BH binaries~\cite{mcmillan}: single
black holes, being heavier than most stars in a globular, sink to the center
via tidal friction, find each other, and make binaries; then the BH/BH binaries
get ``hardened'' (made more compact) by interaction with other black holes,
reaching sizes where gravitational-wave emission will cause them to merge in
less than the age of the universe; and further interactions will often 
eject the BH/BH
binaries from the globular, to interstellar space where they merge.  
Simulations by McMillan and Portegies Zwart~\cite{mcmillan} suggest 
that each dense cluster will make a number of 
such BH/BH binaries, and extrapolations into the universe predict an 
event rate in WB advanced IFOs between $\sim 300$ and $\sim 4000$/yr ---
though the uncertainties are probably larger than suggested by these 
numbers from the literature~\cite{mcmillan,vkalogera}.  

These event rates are very encouraging.  They make it seem quite likely that
the advanced IFOs will observe tens to thousands of BH and NS inspirals per 
year, while
the initial IFOs will be lucky to observe $\sim 1$ per year.  

The observed inspiral waves will last for between $\sim 1000$ and $10,000$
cycles depending on the binary's masses, and will carry detailed
information about the binary and about general relativistic deviations from
Newtonian gravity.  This information can be extracted with good precision using
the method of matched filters. 
Specifically (denoting by $M=M_1+M_2$ the binary's total mass and $\mu =
M_1M_2/M$ its reduced mass):~\cite{cutler_flanagan,poisson_will}

(i) The binary's chirp mass $M_c \equiv \mu^{3/5}M^{2/5}$ 
will typically be measured, from the Newtonian part of the signal's
upward frequency
sweep, to $\sim 0.04\%$ for a NS/NS binary and
$\sim 0.3\%$ for a system containing at least one BH.
(ii) {\it If} we
are confident (e.g., on a statistical basis from measurements of many
previous binaries) that the binary's spins are a few percent or less
of the maximum physically allowed, then the reduced mass $\mu$
will be measured to
$\sim 0.5\%$ for NS/NS and NS/BH binaries, and
$\sim 2\%$ for BH/BH binaries.  (iii) Because the
frequency dependences of the (relativistic) $\mu$ effects
and spin effects are not sufficiently different
to give a clean separation between $\mu$ and the spins,
if we have no prior knowledge of the spins, then
the spin$/\mu$ correlation will
worsen the typical accuracy of $\mu$ by a large factor,
to $\sim 50\%$ for NS/NS, $\sim 90\%$ for NS/BH, and
a factor $\sim 1.5$ for BH/BH.
These worsened accuracies should be improved significantly
(though we do not yet know how much) 
by waveform modulations due to 
spin-induced precession of the orbit~\cite{precess,kidder,apostolatos},
and even without modulational information, a certain
combination of $\mu$ and the spins
will be determined to a few per cent.  (iv) The distance to the
binary (``luminosity distance'' at cosmological distances) can
be inferred, from the observed waveforms, to a precision $\sim 3/\rho 
\alt 30$\%, where $\rho\equiv S/N$ 
is the amplitude signal-to-noise ratio in the total LIGO network
(which must exceed about 8 in order that the false alarm rate be less than the
threshold for detection).  (v) With the aid of 
VIRGO, the location of the binary on the 
sky can be inferred, by time of flight between the detector sites, to a 
precision of order five degrees~\cite{Jaranowski}.
(This angular resolution is rather worse than one might expect from a simple
estimate based on time-of-flight alone.
The reason is that because two LIGO detectors are nearly 
parallel, the addition of VIRGO to the network still
leaves a near-degeneracy between polarization and time-of-flight information.
The addition of a fourth detector with comparable sensitivity in
Australia would decrease the error box on the sky $\Delta \Omega$  by a factor $\sim 4$~\cite{Jaranowski}.)

Advanced LIGO will likely produce a catalog of hundreds or thousands of
binary inspirals and their inferred parameters; this catalog
will be a valuable data base for observational astronomy and cosmology.

Important examples of the general relativistic effects that 
can be detected and measured with precision, in the inspiral
waves, are these: 
(i) As the waves emerge from the binary, some of them get
backscattered one or more times off the binary's spacetime curvature,
producing wave {\it tails}.  These tails act back on the binary,
modifying its radiation reaction force and thence its
inspiral rate in a measurable way.~\cite{poisson_tail}  (ii)  If the orbital
plane is inclined to one or both of the binary's spins, then
the spins drag inertial frames in the binary's
vicinity (the ``Lense-Thirring effect''), this frame dragging causes
the orbit to precess, and the precession modulates the
waveforms~\cite{precess}.  This precession and modulation should
be very strong in a significant fraction of NS/BH binaries~\cite{kalogera1}. 
\subsection{Tidal Disruption of a NS by a BH: Measuring the Nuclear Equation of
State}
\label{sec:NSTidalDisrupt}
As a NS/BH binary spirals inward, its NS experiences ever
increasing tidal forces from the BH's gravitational field (its spacetime
curvature).  In many cases these tidal forces may tear the NS apart before
it begins its final, quick plunge into the hole's horizon.  The gravitational
waves from this tidal disruption and from the termination of inspiral should
carry detailed information about the NS's equation of state (the equation
of state of bulk nuclear matter at $\sim 10$ times the density of an atomic
nucleus).  The disruption waves lie largely in the frequency band 
$\sim 300 {\rm Hz} \alt f \alt 1000$ Hz,
where the wide-band, advanced IFOs have good sensitivity, 
and where narrow-band 
or some other optimized IFO can do even better.  This suggests
that the advanced IFOs may be able to extract new information about the 
nuclear equation
of state from the tidal-disruption waves.  A first, crude estimate~\cite{vallisneri} suggests, for example, that for NS/BH binaries at 140 Mpc 
distance (where the event rate could be a few per year; see the table in Box 
1), tidal-disruption observations may enable the NS
radius $R$ to be measured to a precision $\sim 15\%$, by contrast with its
present uncertainty (for fixed NS mass) of about a factor 2. From the 
measured radii would follow the desired equation-of-state information.
Detailed numerical-relativity simulations will be required to firm up
this estimate, and will be essential as a foundation for interpreting any
tidal-disruption waves that are observed.

The merger waves from NS/NS binaries, by contrast with NS/BH, are likely to
lie outside the band of good advanced-IFO sensitivity --- at frequencies $f\agt
1500$ Hz. However the onset of NS/NS merger, triggered by a plunge of the 
two NS's toward each other, may produce a strong ``cliff'' in the 
waves' spectrum, in a range $f \sim 400$ --- 1000 Hz of good 
sensitivity, and by measuring the cliff frequency we may 
strongly constrain the NS radii (assuming the NS masses 
have been determined accurately from the earlier 
portion of the waveform), and hence also learn about the
nuclear equation of state~\cite{faber}.

\subsection{BH/BH Mergers and Ringdown: 
Observing the Nonlinear Dynamics of Spacetime Curvature}
\label{sec:BBHMerger}

A BH/BH binary evolves through three epochs: a gradual, adiabatic  
inspiral driven by gravitational radiation reaction (discussed in
Sec.\ \ref{sec:nsnsinspiral} above); a merger epoch, during which the
holes' spacetime geometry undergoes violent, highly
nonlinear oscillations; and a ringdown epoch, during which the merged  
black hole undergoes damped pulsations as it settles down into its final,
Kerr-metric, quiescent state.  The inspiral waves are rather well understood
(Sec.\ \ref{sec:nsnsinspiral}), as are the ringdown waves; but the
merger dynamics and waves are hardly 
understood at all.  

It is likely that
in many cases the holes are spinning rapidly and the    
system's three angular momentum vectors (two spins and one orbital)    
are substantially misaligned.  In such cases, the merger dynamics
might be quite rich, with spin-spin and spin-orbit coupling perhaps
strong enough to produce spin flips and complex contortions and
flows of the spacetime geometry as the two holes convert themselves into one.

\begin{figure}
\center{\epsfxsize=4in\epsfbox{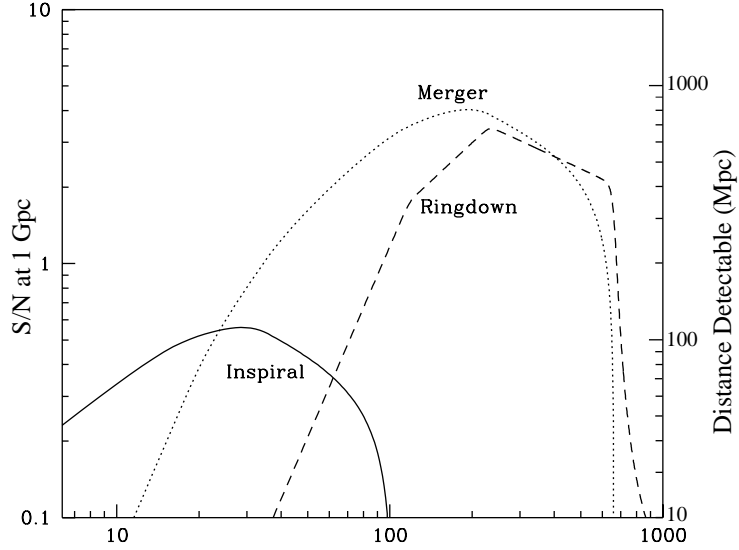}}
\caption{The inspiral, merger, and ringdown waves from equal-mass black-hole
binaries as observed by LIGO's initial interferometers: The 
distance to which
the waves are detectable (right axis) and the signal-to-noise ratio for a
binary at 1Gpc (left axis), as functions of the binary's 
total mass (bottom axis).  (Figure adapted from Flanagan and Hughes~\protect\cite{hughes_flanagan}.)
}
\label{fig:ligo1bh}
\end{figure}

\begin{figure}
\epsfxsize=4.5in\epsfbox{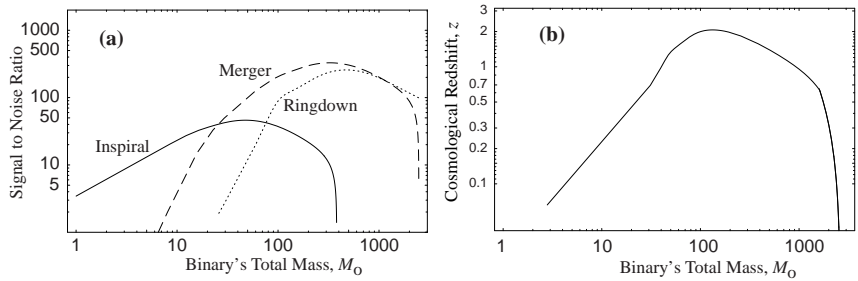} 
\caption{(a) The signal-to-noise ratio for the inspiral, merger and ringdown
waves from fast-spinning, equal-mass black-hole binaries at a fiducial
distance of 1Gpc, as observed by
advanced (LIGO-II) interferometers.  (b) The cosmological redshift out to
which these waves can be seen, assuming a cosmology with
Hubble expansion rate
$H_o = 65$ km/s/Mpc, cold-matter density 0.4 of that required to
close the universe $\Omega_M = 0.4$, and vacuum energy density (cosmological
constant) 0.6 of closure, $\Omega_\Lambda = 0.6$.
(Figure from Ref.~\protect\cite{thorne_LIGOII}, based on the 
LIGO-II wide-band noise curve in Fig.~\protect\ref{fig:SourceSensitivity}   
and the analysis of Flanagan and Hughes~\protect\cite{hughes_flanagan}.)
}
\label{fig:ligoadvbh}
\end{figure}

Flanagan and Hughes~\cite{hughes_flanagan,hughes_flanagan2}
have made plausible estimates about the unknown aspects of the waves
from each epoch --- inspiral, merger and ringdown --- and have used these
estimates to evaluate the signal-to-noise ratios in LIGO and LISA
for equal-mass, fast-spinning BH binaries.  Figures \ref{fig:ligo1bh} and
\ref{fig:ligoadvbh} show their results for the initial (LIGO-I) and
advanced (LIGO-II) IFOs, and Fig.\ \ref{fig:lisabh} (Sec.\ \ref{sec:smbhmerger}
below) shows them for LISA.  [The LIGO-II results in Fig.\ \ref{fig:ligoadvbh}
are updates~\cite{{thorne_LIGOII}} of the original Flanagan-Hughes results,
taking account of the currently planned noise curve for LIGO-II, Fig.\
\ref{fig:SourceSensitivity}.]

Figures \ref{fig:ligo1bh} and \ref{fig:ligoadvbh} suggest that, for $M\alt 24 M_\odot$ (two
$12M_\odot$ holes), the inspiral waves produce most of the signal in LIGO; for
$24M_\odot \alt M \alt 200M_\odot$ the merger waves dominate; and for $200M_\odot
\alt M \alt 1000 M_\odot$, the merger and ringdown waves have roughly equal 
strength.  (We say ``suggest'' because the plausible assumptions made by Flanagan
and Hughes are far from secure; numerical-relativity simulations and/or observations
are needed to firm them up.)  For $M=20M_\odot$ (two $10M_\odot$ holes), 
LIGO's initial and advanced IFOs can detect the waves out to 100Mpc and $z=0.4$
respectively, at which distance population
synthesis calculations suggest event rates $\alt 1/$year and $\sim 30$ to $\sim 4000/$year
respectively (Box 1 and Sec.\ \ref{sec:nsnsinspiral}
above).  Black-hole binaries more massive than $M \sim 50M_\odot$ (two $25M_\odot$ holes)
are unlikely to form
from main-sequence progenitors, but could well form in dense star clusters.  The event
rates at the distances seen by the initial IFOs ($\sim 200$ to $800$ Mpc for masses up
to $\sim 1000 M_\odot$) and by the advanced IFOs (redshifts $z\sim 1$ to $2$ for masses
up to $\sim 2000 M_\odot$)
are exceedingly uncertain; but for the advanced IFOs they could be large.

\subsection{Supernovae and Accretion-Induced Collapse of 
White Dwarfs}
\label{sec:Supernovae}

Type-II supernovae are triggered by the violent collapse of a stellar core
to form a NS or BH.  Despite decades of extensive modeling, we still
understand only poorly the details of the collapse and how it triggers
the explosion, and the birth throes of the NS or BH.
The optical display brings little information, as the stellar material is 
so optically thick that the display is produced hours after the collapse and
from radii $\sim 10,000$ times larger than the NS or BH.  Neutrinos
bring information about the core's temperature during the first
few seconds after the collapse, and gravitational waves could reveal 
the nonspherical dynamics of the core's mass distribution on millisecond timescales
and longer, and the nature of the core's final state: BH or NS. 

Numerical models of supernovae
suggest that, even if it is slowly rotating, a newborn
proto-neutron-star will be convectively unstable, and that the 
gravitational waves
from the convective overturn in the first $\sim$one second of the star's life
may be detectable throughout our galaxy and its orbiting companions, the
Magellanic Clouds~\cite{nsconvection}.  Although the supernova rate 
is low in our galaxy and its companions ($\alt 1/30$yrs), one observed
event could be very valuable scientifically:  The bulk of the supernova's
neutrinos are thought to come from the same convecting material as produces
the gravitational waves, so there should be correlations between the neutrinos
and the waves, which could teach us much about the proto-neutron-star's 
dynamics.

Observations show that many type-II supernovae produce neutron stars and
give them kicks of magnitude as large as $\sim 1000$ km/s.  Indeed, about
half of all radio pulsars are born with a kick larger than 
500 km/s~\cite{cordes}.
The observed kicks suggest that at least some newborn proto-neutron-stars may
be strongly asymmetric, perhaps due to fast rotation, and therefore could
produce significant gravitational radiation.  Rapidly rotating 
proto-neutron-stars
may also be produced by the accretion-induced collapse of white 
dwarf stars (AIC) in some cases, depending on the white-dwarf composition,
central density and accretion rate~\cite{aic}.

If the newborn proto-neutron-star in an AIC or supernova
is spinning so fast that it hangs up
centrifugally at a radius large compared to that of the final NS,
then it may be dynamically (or at least secularly) unstable to 
deforming into a bar-shaped object that tumbles end-over-end, emitting
gravitational waves in LIGO's band of good sensitivity~\cite{shibata_shapiro}.  Recent 
simulations suggest
that the bar will be long-lived, rather than just wrapping itself up
into an axisymmetric shape and disappearing~\cite{barsimulations,barsimulations2}.  The
simulations show that the waves from such a bar may sweep upward in frequency,
due to a gradual shrinkage of the proto-neutron-star, or may sweep downward
due to development of ``Dedekind-like'' internal circulation~\cite{laishapiro}.  
The discovery of the waves from such a proto-neutron-star
and observations
of their frequency evolutions (which should mirror the bar's evolution) would
teach us much.  Though the strengths of the waves and the best
signal processing techniques and thresholds are all 
ill-understood, a rough estimate~\cite{barmoderange}
suggests that the advanced IFOs' range for detection might be 
the distance of the VIRGO cluster of galaxies (about 15 Mpc) and conceivably
significantly larger, suggesting event rates that could be some per year but 
might be far less.

If the advanced WB IFOs detect no gravitational waves from a supernova 
at distance $r$, during a time $T$ preceeding the beginning of the optical
outburst or over a time $T$ during a neutrino outburst, then one can thereby
place a limit on the emitted gravitational-wave energy.   The limit
(as devised by Finn\cite{finn_sn}) is based on the statistics of a cross correlation
of the outputs of the two 4km LIGO interferometers, over the chosen time $T$
and over some frequency band near the minimum of the advanced LIGO
noise curve.
In the band from $\sim 100$ to 300 Hz this limit is 
$ \Delta E_{\rm GW} \alt 0.05 
({r/ 15 {\rm Mpc}})^{2} ({T/ 1 {\rm h}})^{1/2} M_\odot$.  
In the case of a neutrino-emitting supernova in our own galaxy, with $T \sim 1$ sec,
this limit is very impressive: $\sim 10^{-9}M_{\odot}$. 

\subsection{Spinning Neutron Stars}
\label{sec:NSspin}

As a newborn neutron star settles down into its final state, its crust 
solidifies (crystalizes). The solid
crust will assume nearly the oblate axisymmetric shape that 
centrifugal forces are trying to maintain. 
Thereafter the crust has a preferred shape (oblate about
some preferred axis).
One can define the NS's {\it deformation ellipticity} $\epsilon_d$
as the residual ellipticity the NS would then maintain (due to induced crustal
shear stresses) if the NS were spun down to zero frequency without
the crust breaking or otherwise 
relaxing~\cite{cutler_jones,cutleretal_in_prep}:
\begin{equation}
\label{epsilond}
\epsilon_d \sim 
7\times 10^{-8} \bigl(\frac{\nu_{\rm rel}}{\rm kHz} \bigr)^2
\end{equation}
\noindent where
$\nu_{\rm rel}$ is the spin frequency for which the crust is (most) relaxed. 
Again, this is the level of ellipticity that is ``frozen into''
the crust, as opposed to the ellipticity that arises from
centrifugal forces and so ``follows'' the star's spin .
If the star's angular momentum $\vec J$ deviates from the
crust's preferred symmetry axis, the NS will wobble as it spins, with 
small ``wobble angle'' $\theta_w$. The NS will emit GW's with
$h \propto  \theta_w \epsilon_d$
at frequency 
$f_{GW}=f_{\rm rot} + f_{\rm prec}$, where the {\it body frame precession 
frequency} $f_{\rm prec}$ [$ = \epsilon_d f_{\rm rot}(I_{\rm crust}/I)$] is orders of magnitude smaller than $f_{\rm rot}$;
thus $f_{GW}$ is just slightly higher 
than $f_{\rm rot}$~\cite{zimmermann,cutler_jones}.
The NS may also deviate slightly from axisymmetry about its principal axis; 
i.e., it may have a slight ellipticity $\epsilon_e \equiv (I_1-I_2)/I_3$ 
in its equatorial plane.
It will then emit GW's with $h\propto \epsilon_e$  
at twice the rotation frequency, $f_{GW}=2f_{\rm rot}$~\cite{zimmermann}.
The gravitational ellipticity $\epsilon$ referred to in Fig.\ 
\ref{fig:SourceSensitivity} and in Sec.\ \ref{sec:HFOverview} is 
$\epsilon = \epsilon_e$ for deformations around the principal axis, and
$\epsilon = 2\theta_w\epsilon_d$ for deformatins in which $\vec J$ is
inclined to the preferred symmetry axis, producing wobble. 

How large will be $\epsilon_e$ and $\theta_w \epsilon_d$?
Ushomirsky et al.~\cite{ucb} have shown that, for typical NS parameters,
$\epsilon_e < 5\times 10^{-7}(\sigma_{max}/10^{-2})$, where
$\sigma_{max}$ is the (highly uncertain) crustal yield strain, estimated
to be somewhere in the range $10^{-5} < \sigma_{max} < 10^{-1}$. 
Thus rough upper limits are $\epsilon_e < 10^{-5}$ and 
[from Eq.~\ref{epsilond}] $\theta_w \epsilon_d < 10^{-7}$. 
These upper limits are for ellipticities that are
basically supported by crustal shear stresses, but 
there are other possible sources of ellipticity.  For example,
inside the superconducting cores of
many neutron stars there may be trapped magnetic fields with mean
strength $B_{\rm core}\sim10^{13}$G or even
$10^{\rm 15}$G. 
Because such a field is actually concentrated in flux
tubes with $B = B_{\rm crit} \sim 6\times 10^{14}$G surrounded by
field-free superconductor, its mean pressure is $p_B = B_{\rm core} B_{\rm
crit}/8\pi$.  This pressure could produce ellipticity  
$\epsilon_{\rm e} \sim \epsilon_d \sim p_B/p \sim 10^{-6}B_{\rm
core}/10^{15}$G (where $p$ is the core's material pressure). 
We are extremely ignorant, and
correspondingly there is much to be learned from searches for
gravitational waves from spinning neutron stars.

From this discussion we see that the range of plausible gravitational
ellipticities for a spinning neutron star is $\epsilon \alt 10^{-6}$,
though they may get as large as $10^{-5}$.  The strengths of the waves
produced by such ellipticities are shown in Fig.\  \ref{fig:SourceSensitivity}
and the waves' detectability is discussed in Sec.\  \ref{sec:HFOverview}.

Neutron stars gradually spin down, due in part to gravitational-wave
emission but perhaps more strongly due to electromagnetic torques associated
with their spinning magnetic fields and pulsar emission. 
This spin-down reduces the strength of centrifugal forces, and thereby
causes the star's oblateness to decrease, with
an accompanying breakage and resolidification of its crust's crystal structure
(a ``starquake'')~\cite{starquake}.  
In each starquake, $\theta_w$, $\epsilon_e$, and
$\epsilon_d$ will all change suddenly, thereby changing the amplitudes and
frequencies of the
star's two gravitational ``spectral lines'' $f=2f_{\rm rot}$ and
$f=f_{\rm rot} + f_{\rm prec}$.  After each quake, there should be a
healing period in which the star's fluid core and solid crust, now rotating
at different speeds, gradually regain synchronism.
By monitoring the 
amplitudes, frequencies, and phases of the two gravitational-wave
spectral lines, and by 
comparing with timing of
the electromagnetic pulsar emission, one might learn much about the 
physics of the neutron-star interior.

A spinning NS can radiate not only due to built-in asymmetries, as 
discussed above, but also due to a
gravitational-radiation-reaction-driven instability first discovered by
Chandrasekhar~\cite{cfs_chandra} and elucidated in greater detail by 
Friedman and Schutz~\cite{cfs_friedman_schutz}.  In this ``CFS instability'',
density waves travel around the
star in the opposite direction to its rotation, but are dragged forward
by the rotation.  These density waves produce gravitational waves that 
carry positive energy as seen by observers far from the star, but
negative energy from the star's viewpoint; and because the
star thinks it is losing negative energy, its density waves get
amplified.  This intriguing mechanism is similar to that by which
spiral density waves are produced in galaxies.  Although the CFS
instability was once thought ubiquitous for spinning stars~\cite{cfs_friedman_schutz,wagoner}, we now
know that neutron-star viscosity will kill it, stabilizing the star and
turning off the waves, when the star's temperature is above some
limit $\sim 10^{10}{\rm K}$~\cite{cfs_lindblom}
and below some limit $\sim 10^9 {\rm K}$~\cite{cfs_mendell_lindblom}; and correspondingly, the instability
should operate only during the first few years of a neutron
star's life, when $10^9 {\rm K} \alt T \alt 10^{10}\rm K$, and only
for NS spins greater than $\sim 90\%$
of the upper limit set by mass-shedding at the equator~\cite{lindblom_95}.

Shortly after GR15 there was a surge of renewed interest in the CFS instability
when Andersson~\cite{andersson_98} and 
Friedman and Morsink~\cite{friedman_morsink_98} showed that r-modes (driven by Coriolis forces) 
are unstable at all spin values 
(again, in the absence of viscosity), 
and then Lindblom et al.~\cite{lindblom_owen_morsink_98} and 
Andersson et al.~\cite{andersson_etal_99} estimated that in a newborn NS, 
the instability would {\it not} be damped by 
shear or bulk viscosity, but instead would rapidly spin 
down the NS to $\nu \alt 100$ Hz. 
However, since then, a host of other damping mechanisms have 
been suggested, including: (i) wind-up of 
magnetic fields~\cite{rezzolla_2001b}, (ii) dissipation at the 
crust-core bondary~\cite{bildsten_ushomirsky}, 
(iii) bulk viscous damping from weak interactions involving hyperons
in the inner core, which is predicted to kill the instability unless the star is
less masive than about $1.2M_\odot$~\cite{jones,lindblom_owen_2001}; and (iv) 
nonlinear mode-mode coupling, specifically a 3-mode interaction, which 
is predicted to damp out r-modes at a 
maximum dimensionless amplitude $\sim 10^{-3}(\hbox{spin frequency}/1000\hbox{Hz})^{5/2}$~\cite{flanagan_2002}. It therefore appears that
r-modes will not be significant GW sources, after all.

\subsection{Gamma Ray Bursts}
\label{sec:GammaBurst}
Cosmic gamma-ray bursts, observed $\sim$ once per day by detectors on
spacecraft, are thought to be produced by extremely hot shocks in a relativistic fireball,
which is ignited by rapid inflow of gas onto a newborn 
black hole~\cite{gammaburst}.   The fireball is so dense that electromagnetic observations
cannot probe the BH's vicinity;  the observed gamma rays are produced at
radii  $
\sim 10^{13}$ to $10^{15}$ cm --- millions to billions of times larger than the hole itself.
The only hope for probing the fireball's trigger is via gravitational waves:   The BH and its accretion flow are thought
to form violently, by the collapse of a massive star (perhaps initiated by
merger with a companion) [``a hypernova''], or by the merger of a binary
made of compact objects: a NS/NS binary, a BH/NS binary, a BH/white-dwarf
or a BH/He-core binary~\cite{gammaburst}. 
Each of these gamma-burst triggers should emit strong gravitational 
waves that carry detailed information about its source.  

The distribution of $\gamma$-ray bursts is bimodal: there are {\it long 
bursts} (duration $\agt 2$ s) and {\it short bursts} ($\alt 2$ s).
Beppo-Sax has detected afterglows from several long bursts, allowing
their redshifts to be measured. No redshifts are known for short bursts,
but their distance can be estimated from the $\langle V/V_{max}\rangle$ 
statistic~\cite{schmidt}
(assuming a standard profile, and that their event rate tracks the 
star formation rate). The best event rate estimates (by M. Schmidt~\cite{schmidt}) currently 
are: $1/$yr
to $D =  170 ({\theta_0}/{0.1 })^{2/3}$ Mpc for long bursts, and
$1/$yr to D = $250 ({\theta_0}/{0.1 })^{2/3}$ Mpc for short bursts. 
Here the gamma-rays are assumed beamed into solid angle 
$2\pi \theta^2_0$. Clearly, these rates (for $\theta_0 =0.1$) are quite consistent with the estimated NS-NS merger rates 
given in Box 1, and marginally consistent with the BH-NS rate.

If BH-NS mergers are indeed responsible for either long or short bursts,
the (WB) LIGO II detection rate would be $\sim 15-50/$yr.
Of course, these closest sources will generallly not be seen in 
$\gamma$-rays, since they are typically beamed away from us;
the closest burst for which we see the $\gamma$-rays 
(in 1 yr) is at $D = 1.0 (1.5)$ Gpc for long (short) bursts.
Could these be detected in coincidence with GW inspiral events?
Box 1 states that LIGO II (WB) could detect BH-NS (NS-NS)
inspirals to 0.65 (0.30) Gpc. But those distances assume
a random orientation of the source, and assume that $S/N > 8$ is
required for confident detection.
If the $\gamma$-rays are beamed perpendicular to the orbital plane, then
in cases where we detect the burst, we are also at optimal 
``viewing angle'' for the GW's. The amplitude $h$ of the GW's
beamed ``up the axis'' are $\sqrt{5\over2} \sim 1.58$ larger than the rms
value (averaged over the sky, as seen from the source).
Also, the near-coincident detection of a $\gamma$-ray burst in time 
(within several seconds of each other) and on a consistent 
area of the sky, effectively reduces by another factor 
$\sim 1.5$ the GW S/N required for detection with $99\%$ 
confidence~\cite{Kochanek_Piran_93}.
Thus the distances out to which LIGO I and II could detect binary
mergers associated with $\gamma$-ray bursts are 
$\sim 2.4$ times farther than the distances given in Box 1;
for LIGO II: 720 Mpc for NS/NS and 1.55 Gpc for NS/BH mergers.
Thus, if NS/BH mergers
are responsible for long and/or short bursts, with LIGO II one might
expect $\sim 2/$yr simultaneous detections in both GW's and
$\gamma$-rays. If NS/NS mergers are responsible, one might expect
to wait $\sim 5$ yr.

\indent M. van Putten\cite{vanPutten} points out that in the 
above $\gamma$-ray burst scenarios, the accretion torus itself might 
develop large non-axisymmetries and so be a strong source of GW's. 
He develops a picture of long bursts as due to ``suspended'' accretion
around a rapidly rotating BH, with most of the BH's angular momentum
getting transferred to the torus via Maxwell stresses (similar to the 
Blandford-Znajek effect). If the torus develops ``lumps'', van Putten
estimates that $\sim 10\%$ of the BH's spin energy could get radiated
away in GW's--so $E_{GW} \sim 0.3 M_\odot (M_{BH}/10 M_\odot)$.
He predicts a sinusoidal waveform with frequency $f \sim 1$--2
kHz that wanders slowly in time --- which is 
simple enough to allow near-optimal data analysis. Would this be detectable?
The rms $S/N$ would then be (taking $z\ll 1$ and approximating
$f$ and $S_h(f)$ as constants over the signal band, for simplicity), 
$${S\over N} = {{\sqrt{2 E_{GW}}}\over{\pi D f^{1/2} h_n(f) }}.$$
\noindent 
yielding $S/N = 1.4$ for 
$E_{GW}= 0.3 M_\odot$, $D= 150$ Mpc, $f = 10^3$Hz, and
$h_n (\equiv \sqrt{5f S_h(f)}) = 
8 \times 10^{-22}$ (LIGO II WB value at 1 kHz). This does not
seem especially promising, but given the uncertainty in 
$E_{GW}$ and $D$ (for the closest source in one year), it
seems worth looking for.  Moreover, one could improve the sensitivity
in LIGO-II by a factor of a few by reshaping the noise curve 
so it is lower at
$f\sim 1$ kHz, at the price of higher noise at lower frequencies
(cf.\ the difference between the NB and WB LIGO-II curves in Fig.\
\ref{fig:SourceSensitivity}).

If gravitational waves are detected from one or more gamma-burst triggers,
the waves will almost certainly reveal the physical nature of the trigger.
Moreover, by comparing the arrival times of the gravitational waves and the
earliest gamma rays, it should be possible to measure the relative propagation
speeds of light and gravitational waves to an accuracy $\sim 1 \hbox{ sec} /
10^{10} \hbox{ yr} \sim 10^{-17}$. 

If no gravitational waves are detected from any individual gamma burst,
then (as Mohanty, Finn and Romano~\cite{finngamma} have shown) 
the correlation between gamma bursts and
gravitational waves might nevertheless be established by statistical 
studies of the
advanced IFOs' gravitational-wave data in narrow time windows preceeding the
gamma bursts. 

\subsection{Sources of Stochastic Background}
\label{sec:Stochastic}
Searches for a GW stochastic background with  Earth-based detectors
begin with the assumptions that 1) the GW background will certainly 
be smaller than the instrumental noise at all frequencies, and
2) the instrumental noise level will not be known sufficiently well
{\emph a priori} that one can simply search for ``excess noise'' in each
instrument. Thus, stochastic background searches with Earth-based detectors
are based on cross-correlating the outputs of two highly independent detectors.
(The two 4 km LIGO detectors at Hanford, Washington and Livingston, Louisiana
give the best sensitivity).

The most plausible sources of a stochastic gravitational-wave background
in LIGO's frequency band
are processes in the very early universe.  The current best limit on the
strength of such waves is $\Omega \alt 6\times 10^{-6}$, where
$\Omega \equiv \int{\Omega(f) d(\ln f)}$ is the 
integrated GW energy density from all frequency bands.
A wave energy larger than this would have caused the universe to
expand too rapidly through the era of primordial nucleosynthesis
(universe age $\sim $ a few minutes),
thereby distorting the universal abundances of light elements away from
their observed values.  
Other known bounds on $\Omega(f)$ are:
1) $h^2_{100} \Omega(f \sim 10^{-8} {\rm Hz}) < 10^{-8}$ from
pulsar timing; and
2)$h^2_{100} \Omega(f) < 7\times 10^{-11} ({f/{H_0}})^{-2}$ for
$3 \times 10^{-18}h_{100}{ \rm Hz} < f < \times 10^{-16}h_{100}{ \rm Hz}$,
from microwave background anisotropy measurements.  Here $h_{100}$ is the 
Hubble constant in units of $100$ km/s/Mpc. 

The LIGO II interferometers would improve
the current limit at $f \sim 40$ Hz by a factor $\sim 10^3$, to 
$\Omega(f \sim 40 {\rm Hz}) \simeq 5\times 10^{-9}$ --- 
an improvement enabling LIGO to test 
a number of current speculations about the very early universe.  
Inflationary models of the early universe predict that vacuum
fluctuations, created in the Planck era when the universe was being born,
should have been parametrically amplified, during the first $\sim 10^{-25}$ sec
of the universe's life, to produce a stochastic gravitational wave background
in the LIGO band.  Unfortunately, if ``standard'' inflation theory is correct,
then the amplified waves are much too weak for LIGO to detect, $\Omega \alt
10^{-15}$~\cite{turner1}.  The most plausible modifications of standard
inflation push $\Omega$ downward from this~\cite{turner1}, but some less 
plausible
modifications push it upward, to the point of detectability~\cite{grishchuk}.  

The first, tentative efforts to combine superstring theory with inflationary
ideas have produced a new description of the very early universe called the
``pre-big-bang model'', in which string effects cause the gravitational-wave
spectrum to rise steeply at high frequencies --- most likely at frequencies 
above LIGO's band, but quite possibly in or below that 
band~\cite{PreBigBang}.  The result could
be waves strong enough for LIGO to detect.  A non-detection would significantly
constrain the pre-big-bang model.  

There are a wide variety of postulated mechanisms that could have produced
strong gravitational waves, with wavelengths of order the horizon size, at
various epochs in the very early universe.  Those waves (if any) produced
at (universe age)$\sim 10^{-25}$ sec, corresponding to (universe temperature
$T) \sim 10^9$ GeV, would have been redshifted into the LIGO band today
and might be detectable.  The temperature (energy) region $\sim 10^9$ GeV is
{\it tera incognita}; LIGO's advanced detectors will provide our first
opportunity for a serious experimental exploration of it.  Among the
speculated wave-generating mechanisms that could operate there, and that
LIGO could constrain (or discover!), are these: 
\begin{enumerate}
\item[$\bullet$]A first-order phase transition in the states of
quantum fields at $T \sim 10^9$ K.  Such a phase transition would nucleate
bubbles of the new phase that expand at near the speed of light and collide to
produce gravitational waves; and their collisions would also generate
turbulence that radiates waves.  If the transition is strongly first order,
the waves would be strong enough for LIGO's advanced IFOs to detect~\cite{phasetransitions}.  

\item[$\bullet$]Goldstone modes (coherent, classical excitations) of scalar
fields that arise in supersymmetric and string theories.  If strongly excited,
these modes will entail coherent flows of energy that radiate gravitational
waves strong enough for detection~\cite{goldstone}. 

\item[$\bullet$]Coherent excitations of our 3+1 dimensional universe, regarded
as a ``brane'' (defect surface) in a higher dimensional universe.  The
excitations could be of a ``radion'' field that controls the size or curvature
of the additional dimensions, or they could be of the location and shape of
our universe's brane in the higher dimensions; in either case, if there is
an equipartition of energy between these excitations, in the very early
universe, and other forms of energy, then the excitations will produce
gravitational waves easily strong enough for detection by LIGO's advanced IFOs. 
LIGO would thereby probe one or two additional dimensions of size or 
curvature length
$\sim 10^{-10}$ -- $10^{-13}$ mm; by contrast, 
LISA's lower-frequency observations would   
probe lengths $\sim 1$ -- $10^{-5}$ mm. If the number of extra dimensions is
larger than 2, the probes reach to much smaller scales~\cite{mesoscopic,mesoscopic2}
  
\end{enumerate}

Cosmic strings (not to be confused with superstrings), produced in the early
universe, were once regarded as candidates for seeding galaxy formation, but
recent cosmological observations have ruled them out as seeds.  
Nevertheless, it remains
possible that a network of vibrating cosmic strings too weak to seed galaxy
formation was formed in the early universe. LIGO can search for the
presence of such a network in two ways:  (i) Via the stochastic background of
gravitational waves that the strings' vibrations produce; this background would
be strong enough for the advanced IFOs to detect if the strings'
mass per unit length is $\agt 10^{-8}$.~\cite{CosmicStringBackground}
(ii) Via occasional, individually detectable,
strong non-Gaussian bursts (``spikes'') of gravitational waves produced
by kinks or cusps along the string~\cite{DamourVilenkin}. 

\noindent Cusps are points on the string that,
at one instant, move at the speed of light; kinks typically 
arise when two strings cross each other and interconnect.
We now briefly discuss the associated GW spikes as analyzed by 
Damour and Vilenkin~\cite{DamourVilenkin}.
These signals have no free parameters, except the overall amplitude and
the arrival time $t_c$: $h(t) \propto |t-t_c|^{1/3}$ for cusps
and $h(t) \propto |t-t_c|^{2/3}$ for kinks. The time spent in the
detector band is $\sim 1/f_c$ where $f_c$ is the center of the
band. 
Matched filter searches should therefore be trivial, with  
coincidence between detectors (and other vetoes) 
used to remove instrumental bursts. 
Cusp bursts are more highly beamed than kink bursts,
have higher amplitude, and are more detectable (the higher amplitude 
more than compensates for the decreased probability that any 
given cusp burst is
beamed in our direction). Cusp bursts should be detectable 
for a large range of string tension $\mu$ not otherwise constrained:
LIGO II should detect cusp bursts for $\mu \agt 10^{-11}$, while
LISA could detect $\mu \agt 10^{-13}$.
Burst searches thus provide a stronger limit on cosmic
strings than searches for a stochastic, Gaussian background.

\indent Each of these cosmological speculations is plausible, though not highly
likely. (For further explication of most of them, we highly recommend
the review by Allen~\cite{allen_review}.) 
Perhaps their greatest value is to remind us of how terribly
ignorant we are of physics and astrophysics in the domain that LIGO's
advanced IFOs will probe.  Our ignorance may well be even greater than that
of the pioneering radio astronomers of the 1930s 
and X-ray astronomers of the 1960s;
and as there, so also here, the first waves to be discovered may well be
from sources that were previously unknown.  LIGO II and its partners 
could bring us a revolution of insights into the universe 
comparable to the revolutions wrought by radio and X-ray astronomy. 

\section{Low-Frequency Sources Sought by LISA}
\label{sec:lfsources}

\subsection{LISA's Noise Curve and Conventions for Wave Strengths}
\label{subsec:lisanoise}

In discussing low-frequency sources, we shall place them in the context of
the sensitivity of LISA, 
the {\it Laser Interferometer Space Antenna}~\cite{cornerstone}.   As is discussed
in Schutz's chapter in this volume~\cite{schutz_gr16}, LISA consists of three
drag-free spacecraft at the corners of a 5-million-kilometer triangle.  The
spacecraft track the distances to each other using laser beams, thereby searching
for gravitational waves in the $10^{-4}$--$1$ Hz low-frequency band.  LISA is
tentatively scheduled for launch in 2011 as a joint European-American mission
with a nominal lifetime of 5 years, though it could well continue to collect GW
data much longer than that.   By the time LISA begins its GW
searches, the second-generation earth-based detectors (e.g.\ the
advanced LIGO-II IFOs) described
in the last section could be nearing the end of their observations, and a transition
to a third generation might be near at hand.

\begin{figure}
\epsfxsize=4.5in\epsfbox{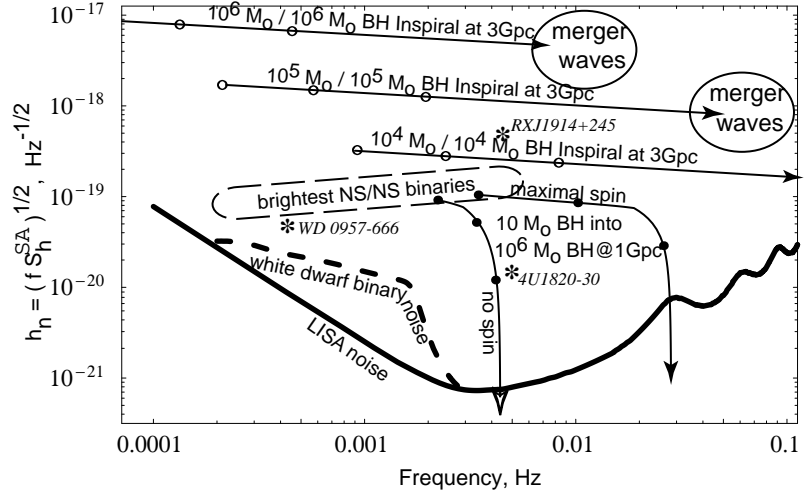}
\caption{LISA's sky- and polarization-averaged rms noise $h_n(f)$ in a 
bandwidth equal to frequency,~\protect\cite{lisa_noise_curve} 
compared with the strengths of the waves from several low-frequency sources. 
The wave strengths are plotted so the height above the noise curve is the 
$S/N$ for a wave search using optimal signal processing; see text for
further detail.  For frequency-sweeping waves (arrowed curves) the dots are,
from left to right, the signal strength and frequency 1 year, 1 month, 
and one day before the end. The thick-dashed curve is a background
of waves from WD-WD binaries that are so numerous they cannot be removed
from the data, and so constitute a noise when searching for other waves.
}
\label{fig:lisa_noise}
\end{figure}

Figure \ref{fig:lisa_noise} depicts LISA's noise and compares it with the wave
strengths for various low-frequency sources, in much the same manner as
Fig.\ \ref{fig:SourceSensitivity} depicted LIGO's noise and compared it with
high-frequency wave strengths.  
However, there are
important differences in conventions
between the LISA Fig.\ \ref{fig:lisa_noise} and the LIGO Fig.\ \ref{fig:SourceSensitivity}:  

{\it First:} {\bf For LIGO} we plotted vertically the square root of the
spectral density of the fractional arm-length difference,
$\tilde h(f) \equiv \sqrt{S_{\Delta L/L}(f)}$ (noise per root Hz) for a
single 4 km IF0. {\bf For LISA}, by contrast, we plot
the rms GW strain noise in a bandwidth equal to frequency $\Delta f = f$,
averaged over the sky and over GW polarizations, 
for the full 3-spacecraft LISA system:  
$h_n = \sqrt{f  S_h^{\rm SA}(f)}$, where $S_h^{\rm SA}(f)$ is the noise spectral density inverse-averaged over the sky and 
polarizations~\cite{finn_thorne}. 
(For LIGO the analogous quantity is $h_n = \sqrt{5f} \tilde h$, where the $\sqrt{5}$
accounts for the average over the sky and polarizations.)
In many discussions of LISA a different quantity
is used to characterize the noise:  
the amplitude sensitivity to periodic
sources for a one-year integration time and a signal-to-noise ratio of 5, 
averaged over the sky and over source polarizations.  That quantity,
$h^{\rm SA}_{\rm SN5,1yr}$, is related to our rms noise $h_n$ by
\begin{equation}
h_n = {\sqrt{f\times 1{\rm yr}}\over 5} 
h^{\rm SA}_{\rm SN5,1yr}\;. 
\label{hn}
\end{equation}

{\it Second:} Our wave-strength convention {\bf for LISA} 
(Fig.\ \ref{fig:lisa_noise}) is also different from that for LIGO (Fig.\ \ref{fig:SourceSensitivity}):  The 
height of a source point or curve above the LISA noise curve is equal to
the signal-to-noise ratio $S/N$ in a search for the source's waves
using optimal signal processing.  By contrast, {\bf for LIGO}
(Fig.\ \ref{fig:SourceSensitivity}), the height above
the noise curve is the signal-to-threshold ratio $S/T$, with the
threshold being that at which the false alarm probability is one per cent 
when using the best currently known, practical data analysis algorithm
(typically $T\agt 5N$).  The difference in convention is motivated
by the fact that we know little as yet about practical LISA
data analysis algorithms and thus don't know with much confidence how to set
thresholds, whereas LIGO data analysis is fairly mature. 

For LISA, when evaluating the $S/N$'s that determine the height of the source
points in
Fig.\ \ref{fig:lisa_noise}, we assume 
for periodic sources (e.g. white dwarf binaries) an integration time
of five years (the nominal mission lifetime), and for sources with a 
large frequency sweep, we assume
at a given frequency $f$ an integration time that is the shorter of 
the time until the signal stops, and the time to sweep through a frequency
band $\Delta f = f$.

Notice that 
LISA's minimum noise $h_n \simeq 10^{-21}$ is
nearly the same as that of LIGO's first interferometers (Fig.\ 
\ref{fig:SourceSensitivity} with $h_n = \sqrt{5f}\tilde h$), but at 
100,000 times lower frequency $f$.  Since
the waves' energy flux scales as $f^2 h^2$, this corresponds to $10^{10}$
better energy sensitivity than LIGO-I.

LISA can detect and study, simultaneously, a wide variety of different
sources scattered over all directions on the sky.  The key to
distinguishing the different sources is the different time evolution of
their waveforms.  The key to determining each source's direction, and
confirming that it is real and not just noise, is the manner in which
its waves' amplitude and frequency are modulated by LISA's complicated
orbital motion---a motion in which the interferometer triangle rotates around
its center and around the normal to the ecliptic plane 
once per year as the spacecraft orbit the sun; see the discussion
by Schutz~\cite{schutz_gr16} elsewhere in this volume.  
Most sources will
be observed for of order a year or longer, thereby making full use of these
modulations.

Our understanding of sources and data analysis for LISA has
evolved considerably since GR15. The principal sources are
(i) short-period stellar-mass binaries --- both galactic and extragalactic,
(ii) the inspiral and merger of massive ($\sim 10^6 M_\odot$) BH binaries,
(iii) the inspiral and capture 
of stellar-mass compact objects (white dwarfs, NS's and BH's) 
into massive BH's, (iv) the gravitational collape of supermassive stars,
and
(v) a stochastic GW background from the early universe. We discuss these in turn
in the following sections.

\subsection{Short-Period Binaries}
\label{sec:binaries}
Unlike the case for ground-based detectors, there are known GW sources
that LISA is certain to detect at its planned sensitivity:
short-period binary stars in our galaxy.  There are currently about a dozen
galactic binaries with GW frequencies above 0.1 mHz that LISA
would be able to detect and study in detail.  These 
include:~\cite{phinney_lisa_scireq}
(i) White-Dwarf / white-dwarf (WD/WD) binaries; the most detectable known
example is WD 0957-666 (shown in Fig.\ \ref{fig:lisa_noise}), 
with masses $0.37$ and $0.32M_\odot$, distance
from Earth 100 pc, and time to merger $2\times10^8$ years.
(ii) Am CVn
stars --- a white dwarf (WD) that accretes from a low-mass helium-star 
companion;
the most detectable known example is RXJ1914+245 (shown in Fig.\ 
\ref{fig:lisa_noise}), 
a $0.07M_\odot$ helium star orbiting a $0.6M_\odot$
WD at 100pc distance from Earth.  
(iii) Low-mass X-ray binaries --- a NS that accretes from a low-mass companion;
the most detectable known example is 4U1820-30 (shown in Fig.\ 
\ref{fig:lisa_noise}), a  
$<0.1 M_\odot$ star orbiting a $1.4 M_\odot$ NS at $8100$ pc from Earth.
These sources will provide an important test 
that LISA is functioning as expected (as well as a test of the general relativistic
theory of GW emission, in the weak-gravity limit).

These optically-known binaries are just the tip of the iceberg;
there are an estimated $10^{8-9}$ galactic binaries with GW frequencies $f > 0.1$ mHz
(most are WD/WD). At frequencies below $\sim 2$ mHz, there are
many binaries per resolvable frequency 
bin $\Delta f = 1/T_{obs} \sim 10^{-8}$Hz --- too many to be fit and
removed from LISA's data, so they constitute a source of 
``confusion noise'' that 
will dominate over the instrumental noise at these frequencies.
Thus LISA truly suffers from an ``embarrassment of riches.''
But there is also some science to be extracted from this confusion noise:
because the WD/WD background is anisotropic, the confusion
noise level is modulated by LISA's time-varying antenna pattern on the sky.
From this one can extract with few-percent accuracy
the $m=1$ and $m=2$ moments (in spherical coordinates centered on the Sun) 
of the WD/WD distribution~\cite{Ungarelli_Vecchio_anisotropy}.

Above $\sim 2$ mHz the density of sources thins to less than $\sim 0.5$
per frequency bin, and one expects to be able to 
measure individually all these binaries, about $10^4$ of them, and remove
them from the data.  
The confusion noise therefore 
drops dramatically above $\sim 2$ mHz --- to a level set by the
extragalactic binary background (about $10$ times smaller than the
galactic background) --- and crosses below the instrumental noise curve;
see Fig.~\ref{fig:lisa_noise}.

Although it should be possible to remove the binaries above 2 mHz
from LISA's data so as to search there for other waves (e.g.\ from inspiral
of small mass BH's into supermassive BH's; cf.\ Fig.~\ref{fig:lisa_noise}), 
no effort has yet been put into the task of developing practical
data analysis algorithms to perform the removal.  Designing and optimizing
such algorithms is an important challenge for the next few years. 

For the $\sim 10^3$ highest-frequency binaries, one should be able
to measure the binary's rate of change of frequency $\dot f$ 
(due to inspiral driven by GW emission). From the measured 
$\dot f$ and $f$ one can
infer the binary's chirp mass $M_c$, and from $M_c$ and the measured 
amplitudes of the binary's two polarizations once can infer its inclination
to the line of sight and its distance.
For binaries with $f \sim 10^{-2}$ Hz, one expects to measure
the angular location to $\sim 1^\circ$~\cite {Cutler_Vecchio_mono} 
and the distance to $\sim 2\%$~\cite{hughes_private}.
Thereby one can obtain a 3-D map of the galaxy with distances 
unsullied by problems of dust, etc. 

\subsection{Mergers of Supermassive Black Holes}
\label{sec:smbhmerger}

\begin{figure}
\center{\epsfxsize=4.0in\epsfbox{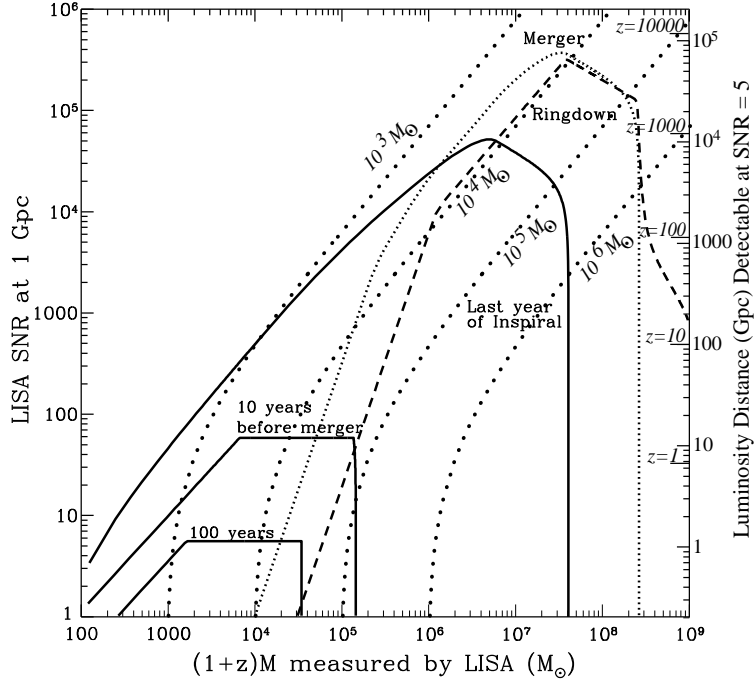}}
\caption{The waves from equal-mass, supermassive black-hole
binaries as observed by LISA in one year of integration time. 
Plotted horizontally is the binary's total mass $M$ multiplied
by 1 plus the binary's cosmological redshift $z$. 
The solid, dashed, and tight-dotted curves refer to the left axis,
which shows the estimated S/N for LISA's observations when the binary
is 1 Gpc from earth (redshift $z\simeq 0.2$).
The wide-spaced dots are curves of constant binary mass $M$, for use with the
right axis, which shows the luminosity distance and redshift 
to which the binary can be detected, with S/N = 5 (look at the intersection
of a wide-dotted curve with a solid, dashed, or tight-dotted curve and then
project horizontally onto the right axis).  (The assumed cosmology is
$H_o=65$ km/s/Mpc, $\Omega_\Lambda = 0.6$, $\Omega_M = 0.4$.)
The bottom-most curves are the
signal strengths after one year of signal integration, for BH/BH binaries 10
years and 100 years before their merger.  
(Figure adapted from Flanagan and Hughes \protect\cite{hughes_flanagan}.)
}
\label{fig:lisabh}
\end{figure}

The merger of two supermassive black holes (SMBHs; 
$10^4- 10^7 M_\odot$) at $z=1$ could be detected by 
LISA with S/N of thousands; 
see Fig.~\ref{fig:lisabh}.
This S/N level is at least 10 times greater than we can expect
with ground-based detectors
(Figs.\ \protect\ref{fig:ligo1bh} and \protect\ref{fig:ligoadvbh}) ---
so high that the inspiral portion of the
waveform could typically yield the binary parameters to the following 
accuracies:
individual masses to 
$\Delta M_i/M_i \sim 10^{-4}$, the so-called spin-orbit parameter $\beta$ to 
$\Delta \beta \sim 10^{-3}$, the distance to 
$\Delta D/D \sim 10^{-3} - 10^{-2}$, and the position of the source
on the sky to $\sim 10^{-4}$ sr~\cite{vecchio99}. 
Knowing the binary's initial conditions with such high accuracy, it
should be possible to make high-precision comparisons of numerical
relativity simulations with the observed merger waveforms --- thereby
extending LIGO's exploration of the nonlinear dynamics of
curved spacetime (Sec.~\ref{sec:BBHMerger})
into a domain of far higher accuracy.  For example,
as the gravitational waves from a binary's merger depart from
their source, the waves' energy should create (via the nonlinearity of 
Einstein's
field equations) a secondary wave called the ``Christodoulou memory''~\cite{christodoulou,thorne_memory,wiseman_will_memory}, which 
LISA should easily be able to measure, while LIGO might not quite see it.  

Unfortunately, it is far from obvious whether the event rate for such 
SMBH mergers will be interestingly high:  By contrast with advanced
LIGO IFOs, where Kalogera and others estimate tens to thousands of BH/BH mergers per year
(Box 1),  Haehnelt~\cite{haehnelt} and others estimate for LISA's SMBH mergers
out to $z=1$ a rate in the
range $10^{-1}-10^{2}$ per year. 
Even if the coalescence rate is only 0.1/yr, then LISA should still see $\sim 3$ BH/BH
binaries with $3000M_\odot \alt M \alt 10^5 M_\odot$
that are $\sim 30$ years away from their final merger.  These slowly
inspiraling binaries should be visible, with one year of integration, out to 
a redshift $z \sim 1$ (bottom part of Fig.\ \ref{fig:lisabh}).

Because the SMBH-merger S/N is so
high at $z=1$, LISA could detect such mergers to redshifts
$z\gg 1$.  SMBH's are not expected to form in the universe at redshifts any
larger than $z\sim 30$, and perhaps somewhat less than this.  Figure
\ref{fig:lisabh} shows that the merger waves can be detected to distances
well beyond $z\sim30$ for masses in the range $M\sim 10^4$ to $\sim 10^7 M_\odot$.
(Below a few times $10^5 M_\odot$ a crucial role in making this possible
is played by the redshift itself, which lowers the waves' frequency,
bringing them into LISA's band of good sensitivity; see Fig.\ 
\ref{fig:lisa_noise}.) 

In the standard picture of bottom-up structure formation, large 
galaxies are formed by the merger of protogalaxies (which are
themselves formed from the mergers of smaller protogalaxies). If the
the merging protogalaxies contain $10^5 M_\odot$ BH's at their centers,
dynamical friction will bring the BH's to the common center, plausibly 
(though not assuredly) to the point where gravitational radiation
takes over and drives the BH's to merger in less than a Hubble time.
Based on this picture, 
Phinney estimates that the merger rate for $10^5 M_\odot$ BH's
would be  $\sim 1500/$yr at $z\sim 7$, $\sim 500/$yr at $z\sim 10$, and 
$\sim 1/$yr at $z\sim 20$~\cite{phinney_lisa_scireq,phinney_private} ---
rates comparable to those of advanced LIGO IFOs (Box 1).
Thus LISA has the possibility to reach back and shed light on the 
earliest epochs of galaxy and SMBH formation

\subsection {Inspirals of $\sim 1-1000 M_\odot$ Compact Objects into Massive BH's}
\label{sec:smbhinspiral}
LISA has the potential to 
observe the final inspiral waves from compact objects (WD's, NS's and BH's) 
of any mass $\mu\agt 1M_\odot$ spiraling into central bodies of mass 
$3\times 10^5 M_\odot \alt M \alt 3\times10^7M_\odot$ out to 3Gpc.  
Figure \ref{fig:lisa_noise} shows the
example of a $10M_\odot$ black hole spiraling into a $10^6M_\odot$ SMBH
at 1 Gpc distance.  The inspiral orbit and waves are strongly influenced
by the SMBH spin.  Two cases are shown~\cite{finn_thorne}: 
an inspiraling circular orbit 
around a non-spinning hole, and a prograde, circular, equatorial orbit 
around a near maximally spinning hole ($a/M = 0.999$ in Kerr-metric 
language).  For each case three dots are shown
on the signal curve --- from left to right the signal strength and
frequency one year, one month, and one day before the small BH plunges
into the horizon of the SMBH.  Table \ref{Table.tbl} shows, at each of
these times, the ratio of the orbital circumference to the SMBH horizon 
circumference, ${\cal C}/{\cal C}_H$, the number of GW cycles left until
plunge, $N_{\rm cyc}$, and the signal to noise in a bandwidth equal to
frequency (or, if that band extends beyond plunge, then the total signal to 
noise in the remainder of the waves), $S/N$. 

\begin{table}
\center{
\caption{The waves from a $10M_\odot$ in circular equatorial orbit around
a $10^6 M_\odot$ SMBH.}
\begin{tabular}{ccrrr}
\hline \hline
SMBH Spin & Time to Plunge & ${\cal C}/{\cal C}_H$ & $N_{\rm cyc}$ & $S/N$ \\ \hline \\
0.999 & 1 yr & 6.80 & 185,000 & 200 \\
0.999 & 1 mo & 3.05 & 40,900 & 60 \\
0.999 & 1 day & 1.30 & 2,320 & 5 \\
0 & 1 yr & 9.46 & 85,000 & 60 \\
0 & 1 mo & 7.14 & 9,650 & 70 \\
0 & 1 day & 6.22 & 366 & 20 \\  \hline \hline
\end{tabular}
\label{Table.tbl}
}
\end{table}

The numbers in Table \ref{Table.tbl}  are remarkable:  The small BH spends its last year
in the immediate vicinity of the SMBH horizon, emitting $\sim 100,000$ cycles of waves.
These waves should carry, encoded in themselves, high-accuracy information 
about the SMBH~\cite{ryan_accuracy}:  

Ryan~\cite{ryanmap}
has shown that,
when any small-mass, compact object spirals into a 
much more massive, 
compact central body, the inspiral waves will carry a ``map'' of the 
massive body's external spacetime geometry.  Since that  
geometry is
uniquely characterized by the values of the body's multiple moments, we can say
equivalently that the inspiral waves carry, encoded in themselves, the values 
of all the body's multipole moments.  
By measuring the inspiral waveforms and extracting their map (i.e., measuring
the lowest few multipole moments), we can determine whether
the massive central body is a black hole or some other kind 
of exotic compact object~\cite{ryanmap}.   The huge number of wave cycles,
in the example in Table \ref{Table.tbl}, implies the possibility of a high-precision map.

While the
S/N values in Table  \ref{Table.tbl} are large enough (up to 200) 
to seem encouraging for detecting
these waves and extracting the map that they carry, there are serious worries.  These S/N's
assume optimal signal processing (via matched filters), but it is highly unlikely that
optimal processing can be achieved.  The reasons are two:

{\it First:} 
Models of galactic nuclei, where massive holes (or other massive central
bodies) reside, suggest that
inspiraling stars and small holes typically will be in rather eccentric
orbits~\cite{hils_bender,sigurdsson_rees}.  
This is because they get injected into such orbits via
gravitational deflections off other stars, and by the time gravitational
radiation reaction becomes the dominant orbital driving force, there is
not enough inspiral left to strongly circularize their orbits.  Such orbital
eccentricity will complicate the waveforms and complicate the extraction
of information from them.  In fact, the waveforms are likely to be so complex
that the number of discrete templates needed in a
in one-year coherent-integration search for these waves may be enormously
larger than modern computers can handle.  As a result, the signal processing
may have to be done coherently over short stretches of time with incoherent
techniques used to combine successive stretches, leading to a significant
loss of $S/N$.  Efforts to generate waveforms from generic orbits and then
scope out this problem are barely getting started, as of
winter 2001--02.  (The waveforms are fairly well understood for generic orbits
in the SMBH equatorial plane~\cite{glampedakis} and for circular orbits inclined
to the equator,~\cite{hughes_circular} but inclined generic orbits have
not yet been studied, and the counting of templates even for equatorial
orbits and inclined circular orbits has barely begun~\cite{list_wg1}.  Although
the influence of radiation reaction on the orbital evolution is not yet
fully understood~\cite{radreact}, it is probably known well enough  to scope
out the data analysis problem.)

{\it Second:} 
There may be a significant 
loss of $S/N$ due to ``source confusion'' --- i.e.\ due
to the problem of separating out individual inspiral waveforms from 
the background due to
other, weaker inspiral sources.  At high frequencies this confusion noise
could perhaps dominate all other noise sources, just as WD/WD confusion noise
dominates at $10^{-4} - 10^{-3}$ Hz (Fig.~\ref{fig:lisa_noise}). Efforts
to scope this out are also barely getting started.

The event rates for inspirals into SMBH's  are not well known. 
Phinney~\cite{phinney_private} 
(relying in part on simulations of star-cluster evolution around a SMBH
by Freitag~\cite{freitag}) 
has shown that the detection rate is likely to be
dominated by $\sim 10 M_\odot$ BH's spiraling into 
$\sim 10^6 M_\odot$ SMBH's. (Phinney shows that 
$\sim 10^6 M_\odot$ SMBH's are likely to have larger
relaxed stellar cusps around them than either smaller 
or bigger BH's; coincidentally $10^6 M_\odot$ is the optimum SMBH mass for
LISA.  And while inspirals of $\sim 10 M_\odot$ BH's probably have a smaller
space density than inspirals of $\sim 1.4 M_\odot$ NS's, because
of their larger S/N, they can be detected throughout a volume that
is $(\sqrt{10/1.4})^3 \sim 20$ times larger.)

The estimated inspiral rate  for
$\sim 10 M_\odot$ BH's is $\sim 1-10/$yr out to 1 Gpc~\cite{phinney_private,sigurdsson_rees}, where the $S/N$'s would be $\sim 60$ to $\sim 200$ {\it if
optimal signal processing were possible}.  If the loss of $S/N$ due to 
signal-processing problems is much larger than an order of magnitude (which
it might be), then there could be difficulty detecting these inspirals.
This motivates exploring a change of
LISA's design so its noise curve is lowered by a factor 4.  Since such a
change might significantly increase LISA's cost,  it is greatly
hoped that a
scoping of the data analysis problem and a firming of the event
rate will reveal that no design change is needed! 

\subsection{Collapse of Supermassive Stars}
\label{sec:smscollapse}

  New and Shapiro~\cite{shapiro} have investigated the possibility of
substantial GW emission from the collapse of supermassive stars (SMS's)
to form SMBH's in galactic nuclei. The formation mechanism
for SMBH's is still highly uncertain; one path could lead
through rotating SMS's. While non-rotating SMS's suffer from
a well-known radial collapse instability (arising from GR effects), even
a small amount of rotation can stabilize the star~\cite{fowler}. 
As the SMS cools and contracts, the ratio
of $\beta = T/|W|$ (ratio of kinetic to gravitational binding energy)
grows, and (for sufficiently large initial $\beta$) 
the star eventually becomes unstable either to mass-shedding
at the equator or to a dynamical, $m=2$ (bar-mode) instability.
New and Shapiro show that if viscosity and magnetic fields are
insufficient to enforce uniform rotation (so that angular momentum
is conserved on cylinders during the contraction), then the bar-mode
instability is likely to occur first. 

If the bar-mode perturbation grows
to order unity and is long-lived, a substantial fraction 
of the star's rest mass-energy could be radiated in GW's. These
waves end up in the LISA band for a range of plausible initial parameters.
For example, for SMS mass $M \sim 10^6 M_\odot$, initial (pre-collapse)
radius $R\sim 10^{17}$cm and initial $\beta \sim 10^{-6}$, 
New and Shapiro~\cite{shapiro} estimate the SMS could radiate away 
$\sim 10^3 M_{\odot}$ in GW's at $f \sim 3\times 10^{-4}$Hz,
in a ``burst'' lasting $\hat\tau \sim 1{\rm yr}$, which is $n =f\hat\tau
\sim 10^4$ cycles.  {\it If} optimal signal processing could be used
to search for these waves, the signal to noise ratio would be $S/N \simeq
h_c/h_n$ where 
$h_c \equiv h \sqrt{n}\sim 10^{-18}$ at 1Gpc distance; here $h$ is the wave amplitude as 
estimated by New and Shapiro (corrected by them from a different value
given in their paper), and $h_c$ is called
the waves' {\it characteristic amplitude}~\cite{300yrs}.  Comparing with
LISA's noise curve (Fig.\ \ref{fig:lisa_noise}) we see that $S/N \sim 40$
for optimal signal processing.  With sufficiently accurate templates from
numerical simulations of the collapse, it might be possible to achieve 
a search within a factor 4 of optimal, so $S/N \sim 10$, which would be 
adequate for confident detection; but as yet it is far from clear one can do so well ---
especially as the search must be performed in the thick of the WD-WD
background.  To get a clearer handle on the prospects will require 
considerable analysis --- both source simulations and data-analysis 
explorations.

\subsection{LISA and the GW Stochastic Background}
\label{sec:lisabackground}

While ground-based detectors can search for a
stochastic background by cross-correlating their outputs, this
method is not available to LISA, since only one space-based
detector is currently planned.

However, it is possible to combine the
signals from the 3 spacecraft to construct a ``symmetrized Sagnac observable'',
which is practically insensitive to both laser phase noise and 
gravitational waves~\cite{lisa_noise_curve}. With the aid of this observable,
one can directly measure 
(symmetrized averages of) the proof-mass noise
and optical-path noise, and thereby determine the instrumental
noise power in the (sum of the) 3 Michelson channels used to detect
GW's. 
Any
excess ``noise'' above this power is due to GW's. 
This method is most effective in frequency bands where either the proof-mass 
noise or optical-path noise greatly dominates over
the other. With this approach, in a one-year integration 
LISA could detect $h^2_{100}\Omega \agt 10^{-11}$ in
either the $10^{-4.5} < f < 10^{-3.5}{ \rm Hz}$ 
or $10^{-2.8} <  f < 10^{-1.6}{ \rm Hz}$ bands~\cite{bender_hogan}.
Unfortunately, 
LISA's sensitivity to a GW background of {\it early-universe} origin
is about an order-of-magnitude worse than this, since 
any primordial source with $\Omega(f) \alt 10^{-10}$
will be ``covered up'' by (the isotropic portion of)
the galactic and extragalactic WD/WD binary backgrounds.  LISA's low-frequency
(LF) sensitivity of $\Omega \sim 10^{-10}$ is about two orders of magnitude
better than the adjacent-band sensitivities of advanced LIGO IFOs 
($\sim 5\times10^{-9}$ in the HF band) and current pulsar timing
($\sim 10^{-8}$ in the VLF band); Sec.\ \ref{sec:Stochastic}.

In Sec.\ \ref{sec:Stochastic} we discussed a number of speculative
early-universe sources that could produce GWs in earth-based detectors' 
HF band.  Variants of all these sources could also radiate in
LISA's LF band:  (i) Universal inflation (with the standard model predicting
waves much too weak for LISA to see,  $\Omega \alt 10^{-16}$--$10^{-15}$,
while some less plausible models such as the pre-big-bang suggest
stronger waves); (ii) Phase transitions of quantum fields (LISA's
frequency band is ideal for probing the electroweak phase transition,
which should have occurred at universe age $\sim 10^{-12}$ sec and
if strongly first order could have produced waves strong enough for
LISA to see~\cite{turner_kosowsky});  (iii) Goldstone modes of scalar
fields that arise in supersymmetric and string theories;  (iv) excitations
of our universe as a brane in higher dimensions (with LISA probing
excitation lengthscales $\sim 1$--$10^{-5}$ mm if there are one or
two relevant higher dimensions); and (v) Cosmic strings (with LISA's searches
for bursts from string cusps and kinks able to constrain string tensions 
to $\mu \alt 10^{-13}$).  For further discussion and references, see Sec.\ \ref{sec:Stochastic}.

There is some hope that a follow-on mission to LISA might
reach the sensitivity required for probing standard inflation,
$\Omega(f) \sim 10^{-16}-10^{-15}$.  To do so will require that the mission's detectors
operate outside the regime dominated by close binaries.
A ``short-LISA'' with $\sim 100$ times shorter arms would have
good sensitivity at $f\sim 1 { \rm Hz}$, where the WD binaries
disappear and the binary background is dominated by
NS binaries; a constellation of 3 short-LISA's spaced evenly
around the Sun (along Earth's orbit) could have 
angular resolution $\Delta\theta
\sim 10^{-3}$ rad, allowing (probably) 
the NS binaries to be detected individually and subtracted out.
At present, such a constellation of $\sim 9-12$ satellites 
seems the best possibility for probing the inflationary GW
background from space. 

\section{Conclusions}
\label{sec:conclusions}

It is now about 40 years since Joseph Weber initiated his pioneering
development of gravitational-wave detectors~\cite{weber}, and 30 years
since Robert Forward~\cite{forward} and Rainer Weiss~\cite{weiss}
initiated work on interferometric detectors. Since then, hundreds of
talented experimental physicists have struggled to improve the
sensitivities of gravitational-wave detectors, and hundreds of theorists have
explored general relativity's predictions.  

These two parallel efforts are now intimately intertwined and are pushing
toward an era in the not distant future, when measured gravitational waveforms
will be compared with theoretical predictions to learn how many and what kinds
of relativistic objects {\it really} populate our Universe, and how these
relativistic objects 
{\it really} are structured and {\it really}
behave when quiescent, when vibrating, and when colliding.
By about the time of GR-18, if not sooner, we should have some answers.

\section*{Acknowledgments}
Much of this manuscript's Section 2 (high-frequency waves) was written in
close consultation with
Lars Bildsten, Alessandra Buonanno, Craig Hogan, 
Vassiliki Kalogera, Benjamin J.\ Owen, E.\ Sterl Phinney, Thomas A.\ Prince, 
Frederic A.\ Rasio, Stuart L.\ Shapiro, Kenneth A.\ Strain, Greg Ushomirsky,
and Robert V.\ Wagoner.  We warmly thank them for their significant 
contributions to this review. We also thank Sam Finn for especially 
valuable input into our sections on supernovae, pulsars, and gamma ray bursts.
Thorne's contributions to this review were supported in part by NSF grant
PHY-0099568 and NASA grant NAG5-10707. Cutler's work was partly supported
by NASA grant NAG5-4093.

\bibliography{cutlerThorne.bib}

\begin{thebibliography}{100}
\bibitem{2kmToNB}
If, in its advanced (``LIGO-II'') incarnation, the 2km interferometer is
  replaced by a narrow-banded 4km interferometer, then for the most important
  wide-band sources, inspiraling binaries, the 3-IFO threshold will be lowered
  (improved) slightly and correspondingly the distances to which the sources
  can be seen will be increased slightly beyond those in Box 1. If the 2km
  interferometer is replaced by a wide-band 4km interferometer, then the
  threshold will be lowered by a factor 1.16, the binaries' observable
  distances will be increased correspondingly by 1.16, and the predicted event
  rates will go up by $1.16^3 \simeq 1.5$.
\bibitem{CosmologicalExtrapolation}
When extrapolating out to cosmological distances, $z\agt 0.1$, we must take
  account of cosmological effects. We assume a cosmological model with $H_o =
  65$ km/s-Mpc, $\Omega_M = 0.4$, and we take account of the evolution of the
  numbers of massive stars and thence BH/BH binaries using Eq.~(4) of Madau,
  Ref.~\protect\cite{extrap3}; the result is an event rate, inside redshift
  $z$, that scales as $z^3$ for $0<z<4$, aside from an enhancement by a factor
  $\sim 2$ near $z=1$ and a decrement by a factor $\sim 2$ at $z\sim 2.5$ --
  $4$.
\bibitem{barmoderange}
In order to shrink to normal neutron-star size, the proto-neutron-star must get
  rid of its excess angular momentum. In the idealized case that (i) the
  angular momentum is lost solely to gravitational waves and not to
  hydrodynamic or magnetic processes, and (ii) the LIGO data are analyzed by
  optimal signal processing, the integrated signal strength depends primarily
  on the excess angular momentum and not on the ellipticity of the bar, and the
  distance to which the signal can be seen is of order 150Mpc (about half that
  for a NS/NS binary whose mass is twice as large). A factor ten reduction of
  amplitude signal to noise (factor 100 reduction of power signal to noise),
  due to non-optimal signal processing and to loss of angular momentum via
  non-gravitational-wave channels, would still leave the source detectable at
  $\sim 15$ Mpc, the distance of Virgo.
\end{thebibliography}
\end{document}